\documentclass{IEEEtran}
\usepackage{cite}
\usepackage{amsmath,amssymb,amsfonts}
\usepackage{algorithmic}
\usepackage{graphicx}
\usepackage{textcomp}
\usepackage{pgfplots}
\def\BibTeX{{\rm B\kern-.05em{\sc i\kern-.025em b}\kern-.08em
    T\kern-.1667em\lower.7ex\hbox{E}\kern-.125emX}}
    
\usepackage{tikz}
\pgfplotsset{
  /pgfplots/xlabel near ticks/.style={
     /pgfplots/every axis x label/.style={
        at={(ticklabel cs:0.5)},anchor=near ticklabel
     }
  },
  /pgfplots/ylabel near ticks/.style={
     /pgfplots/every axis y label/.style={
        at={(ticklabel cs:0.5)},rotate=90,anchor=near ticklabel}
     }
  }

\def\figuretextsize{\footnotesize}
  
\begin{document}
\title{Equalization Enhanced Phase Noise in Coherent Receivers: DSP-Aware Analysis and Shaped Constellations}
\author{Aymeric Arnould, Amirhossein Ghazisaeidi
	\thanks{A. Arnould and A. Ghazisaeidi are with the Nokia Bell Labs, Paris-Saclay, Nozay 91620, France (e-mail: aymeric.arnould@nokia.com). 
	\copyright 2019 IEEE.  Personal use of this material is permitted.  Permission from IEEE must be obtained for all other uses, in any current or future media, including reprinting/republishing this material for advertising or promotional purposes, creating new collective works, for resale or redistribution to servers or lists, or reuse of any copyrighted component of this work in other works.”}
}

\maketitle

\begin{abstract}
We revisit the analysis of equalization-enhanced phase noise (EEPN) arising in coherent receivers from the interaction between the chromatic dispersion compensation by an electronic equalizer and the phase noise of the local oscillator. Through numerical simulations we highlight EEPN characteristics and investigate its impact on the behavior of the carrier phase recovery algorithm. We show that the blind phase search, which is usually used in practice to recover the carrier phase, partially mitigates the EEPN. We detail a numerical approach to predict the system performance including the phase recovery algorithm, and show that taking into account EEPN characteristics relaxes the constraint on the system laser phase noise given by previous pessimistic analytical models. We present experimental validations of our claims, and address future advanced transoceanic systems using 98 GBd probabilistically shaped QAM formats. 
\end{abstract}


\section{Introduction}
\label{sec:introduction}
It has been shown that the interaction between the electronic dispersion compensation (EDC) filter and the phase noise of the local oscillator (LO) laser in the coherent receiver generates a noise called equalization enhanced phase noise (EEPN)\cite{shieh2008}-\cite{farhoudi2012ofc}. The variance of EEPN at EDC output is computed in\cite{shieh2008}. The EEPN-induced constraints on LO linewidth have been experimentally characterized in \cite{xie2009} on quadrature-phase-shift-keying (QPSK) signals, and simulation results in \cite{fatadin2010} highlighted tighter requirements for higher modulation formats such as 16 and 64QAM. 

Assuming that EEPN is an independent additive white Gaussian noise (AWGN), the EEPN variance was added to that of the amplified spontaneous emission (ASE) noise, and the EEPN system penalty was derived in terms of the optical signal to noise ratio (OSNR) in\cite{kakkar2015}. To limit the impact of EEPN on system design, hardware solutions have been proposed, such as digital coherence enhancement (DCE) technique developed in \cite{colavolpe2011}. It is shown in\cite{farhoudi2012} that the EEPN impairing the QPSK signal can be partially mitigated by the Viterbi-Viterbi phase recovery, so that the previous analytical study overestimates the EEPN impact on system design. The probability density function of EEPN noise is studied in\cite{farhoudi2012ofc}. 

In this work, we revisit the EEPN analysis, carrying out a detailed numerical DSP-aware characterization, including the blind phase search (BPS)\cite{pfau} as the carrier phase recovery (CPR) algorithm, and study the EEPN penalty for emerging probabilistically shaped constellations recently proposed for transoceanic distances\cite{amir2017,omar2018}. We believe this revision is timely, as the current industry trend is towards ever increasing symbol-rates: 100 GBd is now practical. And there is also demands for extremely long distance submarine cables, beyond 15000 km. In this context, accurate analysis of EEPN penalty is important. Even if EEPN arises only from the phase noise of the local oscillator laser, we show that taking into account the transmitter laser phase noise is required, since BPS is included in the analysis, and its optimum moving average filter length depends on both transmitter and receiver phase noise, although only the latter results in EEPN. Note that theoretical DSP-aware analysis of EEPN is prohibitively complex, and experimental characterizations require multiple laser sources with widely different linewidths whereas numerical simulation is an efficient approach to this problem. So we adopt numerical simulation is the main analysis tool in this work. We also present some experimental results to complement our studies. More experimental work is required, and will be presented in future works.

This paper is organized as follows. In section \ref{sec:model}, we introduce the reduced model of coherent optical transmission system used in this work. In section \ref{sec:DSPaware}, the DSP-aware numerical approach used to investigate the EEPN is presented. We also present our available (but not sufficient) experimental characterizations. In section \ref{sec:impact}, numerical results are provided to quantify constraints on laser linewidth in the presence of EEPN for transoceanic systems with large accumulated chromatic dispersion, high baud-rate and advanced shaped constellations. In section V, the paper is concluded.

\section{System model}\label{sec:model}
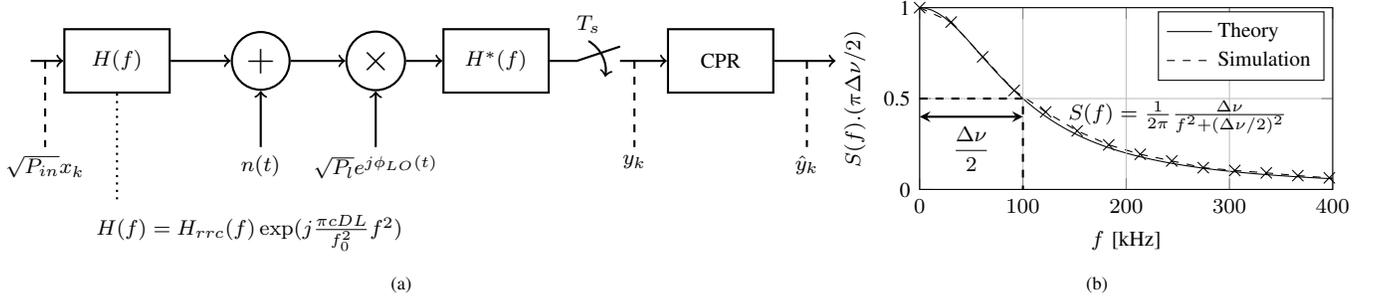
\begin{figure*}[!t]
\begin{minipage}[b]{0.59\textwidth}
\centering
\tikzset{%
  block/.style    = {draw, thick, rectangle, minimum height = 3em,
    minimum width = 1.4cm},
  sum/.style      = {draw, circle, node distance = 1.8cm}, % Adder
  input/.style    = {coordinate}, % Input
  output/.style   = {coordinate} % Output
}
% Defining string as labels of certain blocks.
\newcommand{\suma}{\Large$+$}
{\figuretextsize
\begin{tikzpicture}[scale=\textwidth/14cm, thick, node distance=1.6cm]
%Emitter
\draw
	node at(0,0) [input, name=input] {} 
	node at(1.5,0) [block, align=center] (H1) {$H(f)$}
	;

\draw[->] (input) -- node {} (H1);
\draw[dashed] (0.25,0)--(0.25,-1.5)node[below]{$\sqrt{P_{in}}x_k$};

%DU link
\draw node at (4,0) [sum] (sum1) {\suma};
	
\draw[->] (H1) -- node {} (sum1);
\draw[->] (4,-1.5) node[below]{$n(t)$} --  (sum1);

%Coherent detection
\draw 	node at (6,0) [sum] (prod2) {\Large $\times$}	;
	
\draw[->] (sum1) -- (prod2);
\draw[->] (6,-1.5)node[below]{$\sqrt{P_l}e^{j\phi_{LO}(t)}$} -- (prod2);	

%DSP
\draw
	node [block, align=center,right of=prod2] (H2) {$H^*(f)$}
	node at (12,0) [block, align=center] (CPR) {CPR}
	node at (14,0) [output] (output) {}
	;

\draw[->] (prod2) -- (H2);
\draw (H2) -- (9.5,0) -- (10.25,0.25);
\draw[->] (10.25,0) -- (CPR);
\draw[->] (9.5,0) ++(60:.4) arc(60:-10:.6);
\draw (9.5,0) ++(60:.4) node[above]{$T_s$};
\draw[->] (CPR) -- (output);

\draw[dashed] (10.5,0) -- (10.5,-1.5) node[below]{$y_k$};
\draw[dashed] (13.5,0) -- (13.5,-1.5) node[below]{$\hat{y}_k$};

\draw[dotted] (H1) -- (1.5,-2.5);
\draw (1,-3) node[right]{$H(f) = H_{rrc}(f) \exp(j\frac{\pi c D L}{f_0^2}f^2)$};
\end{tikzpicture}
}
{\scriptsize(a)}
\end{minipage}
\hfill
\begin{minipage}[b]{0.39\textwidth}
\centering
% This file was created by matplotlib2tikz v0.6.18.
\begin{tikzpicture}
\tikzstyle{every node}=[font=\figuretextsize]
\begin{axis}[
axis on top,
legend cell align={left},
legend entries={{Theory},{Simulation}},
tick pos=both,
xlabel={$f$ [kHz]},
ytick={0,0.5,1},
yticklabels={0,0.5,1},
ylabel near ticks,
xlabel near ticks,
xmajorgrids,
xmin=0, xmax=400,
ylabel={$S(f).(\pi\Delta\nu/2)$},
ymajorgrids,
ymin=0, ymax=1,
height=4cm,
width=\textwidth
]
\addlegendimage{no markers, black}
\addlegendimage{no markers, black, dashed}
\addplot [black]
table [row sep=\\]{%
0	1 \\
6.06060606060606	0.996340347666972 \\
12.1212121212121	0.98552036199095 \\
18.1818181818182	0.968 \\
24.2424242424242	0.944492627927146 \\
30.3030303030303	0.915895710681245 \\
36.3636363636364	0.883211678832117 \\
42.4242424242424	0.847470817120622 \\
48.4848484848485	0.809665427509294 \\
54.5454545454546	0.770700636942675 \\
60.6060606060606	0.731363331094694 \\
66.6666666666667	0.692307692307692 \\
72.7272727272727	0.654054054054054 \\
78.7878787878788	0.61699716713881 \\
84.8484848484848	0.581420181526962 \\
90.9090909090909	0.547511312217195 \\
96.969696969697	0.515380974917179 \\
103.030303030303	0.485077951002227 \\
109.090909090909	0.456603773584906 \\
115.151515151515	0.42992499013028 \\
121.212121212121	0.404983265154332 \\
127.272727272727	0.381703470031546 \\
133.333333333333	0.36 \\
139.393939393939	0.33978159126365 \\
145.454545454545	0.320954907161804 \\
151.515151515152	0.303427138478685 \\
157.575757575758	0.287107830213551 \\
163.636363636364	0.271910112359551 \\
169.69696969697	0.257751479289941 \\
175.757575757576	0.24455423310128 \\
181.818181818182	0.232245681381958 \\
187.878787878788	0.22075815933509 \\
193.939393939394	0.210028929604629 \\
200	0.2 \\
206.060606060606	0.190617889025031 \\
212.121212121212	0.181833361162131 \\
218.181818181818	0.173601147776184 \\
224.242424242424	0.165879664889566 \\
230.30303030303	0.158630735615441 \\
236.363636363636	0.151819322459222 \\
242.424242424242	0.145413272800107 \\
248.484848484848	0.139383079482913 \\
254.545454545455	0.133701657458564 \\
260.606060606061	0.128344136711844 \\
266.666666666667	0.123287671232877 \\
272.727272727273	0.118511263467189 \\
278.787878787879	0.113995603475348 \\
284.848484848485	0.109722921914358 \\
290.909090909091	0.105676855895196 \\
296.969696969697	0.101842326755822 \\
303.030303030303	0.0982054288033186 \\
309.090909090909	0.0947533281127643 \\
315.151515151515	0.0914741705165896 \\
321.212121212121	0.0883569979716024 \\
327.272727272727	0.0853916725476358 \\
333.333333333333	0.0825688073394495 \\
339.393939393939	0.0798797036602362 \\
345.454545454545	0.0773162939297124 \\
351.515151515152	0.0748710897215538 \\
357.575757575758	0.0725371344834477 \\
363.636363636364	0.0703079604880883 \\
369.69696969697	0.0681775496149753 \\
375.757575757576	0.0661402976009717 \\
381.818181818182	0.0641909814323608 \\
387.878787878788	0.0623247295827849 \\
393.939393939394	0.060536994830174 \\
400	0.0588235294117647 \\
406.060606060606	0.0571803622998162 \\
412.121212121212	0.0556037784018381 \\
418.181818181818	0.05409029950827 \\
424.242424242424	0.0526366668277829 \\
430.30303030303	0.0512398249658872 \\
436.363636363636	0.0498969072164948 \\
442.424242424242	0.0486052220486499 \\
448.484848484848	0.0473622406819467 \\
454.545454545455	0.0461655856543304 \\
460.606060606061	0.0450130202951267 \\
466.666666666667	0.0439024390243902 \\
472.727272727273	0.0428318584070796 \\
478.787878787879	0.0417994088972479 \\
484.848484848485	0.0408033272134587 \\
490.909090909091	0.0398419492920645 \\
496.969696969697	0.0389137037698767 \\
503.030303030303	0.0380171059521731 \\
509.090909090909	0.0371507522259748 \\
515.151515151515	0.0363133148821234 \\
521.212121212121	0.0355035373129462 \\
527.272727272727	0.0347202295552367 \\
533.333333333333	0.0339622641509434 \\
539.393939393939	0.0332285723003692 \\
545.454545454545	0.0325181402848696 \\
551.515151515152	0.0318300061380177 \\
557.575757575758	0.0311632565460009 \\
563.636363636364	0.0305170239596469 \\
569.69696969697	0.029890483901957 \\
575.757575757576	0.0292828524563715 \\
581.818181818182	0.0286933839222196 \\
587.878787878788	0.0281213686249193 \\
593.939393939394	0.0275661308695102 \\
600	0.027027027027027 \\
};
\addplot [black, dashed, mark=x, mark size=3, mark options={solid}]
table [row sep=\\]{%
0	1 \\
30.517578125	0.920682366370646 \\
61.03515625	0.729445924866694 \\
91.552734375	0.544857734298986 \\
122.0703125	0.4265240294255 \\
152.587890625	0.322324855111213 \\
183.10546875	0.245381493153653 \\
213.623046875	0.193512581687532 \\
244.140625	0.15694693558169 \\
274.658203125	0.119728719482288 \\
305.17578125	0.10439144624477 \\
335.693359375	0.0911162600534845 \\
366.2109375	0.0760312896738163 \\
396.728515625	0.0637338352365426 \\
427.24609375	0.0587623359787917 \\
457.763671875	0.0486952620600972 \\
488.28125	0.0428199738088885 \\
518.798828125	0.0389384663277718 \\
549.31640625	0.0359122858741476 \\
579.833984375	0.0311127898679694 \\
};

\coordinate (a) at (axis cs:0,0.5);
\coordinate (a2) at (axis cs:0,0.4);
\coordinate (b) at (axis cs:100,0.5);
\coordinate (b2) at (axis cs:100,0.4);
\coordinate (c) at (axis cs:100,0);
\coordinate (f) at (axis cs: 250,0.4);
\end{axis}

\draw[dashed,thick] (a) -- (b) -- (c);
\draw[<->,thick,>=stealth] (a2) -- (b2) node[midway,below]{$\dfrac{\Delta\nu}{2}$};
\draw (f) node{$S(f) = \frac{1}{2\pi} \frac{\Delta\nu}{f^2 + (\Delta\nu/2)^2}$};
\end{tikzpicture}
{\scriptsize(b)}
\end{minipage}
\caption{(a) System baseband model with $H(f)$: channel response, $H_{rrc}(f)$: root raised cosine transfer function, roll-off 0.01, $c$: the speed of light, $D$: fiber dispersion coefficient, $L$: transmission length, $f$: frequency, $f_0$: channel center frequency, $H(0)=\sqrt{T_s}$, $T_s$: symbol duration, $n(t)$: ASE noise process, $\phi_{LO}(t)$: local oscillator phase noise process, $P_{in}$: signal mean power, $P_l$: LO power, CPR: carrier phase recovery, $^{*}$: complex conjugation; (b) power spectral density $S(f)$ of the power normalized local oscillator field, $\Delta \nu$: laser linewidth.}
\label{system}
\end{figure*}

We use the equivalent baseband model described in Fig.~\ref{system}a. Previous studies\cite{shieh2008,kakkar2015} have shown that transmitter (TX) laser phase noise impacts only phase recovery but has no influence on EEPN. First, we will omit TX phase noise to characterize EEPN. Then, it will be included in the rest of the study to quantify the impact of EEPN in a realistic system, where phase noise arising from both LO and TX lasers must be considered to evaluate the system performance after DSP. 

A train of symbols $x_k$ at the rate $B=1/T_s$ is passed through the equivalent transfer function $H(f)$ modeling pulse-shaping and fiber chromatic dispersion. The ASE process $n(t)$ is complex circular and additive white Gaussian noise (AWGN). At the receiver, the signal is multiplied by the local oscillator field. The LO phase noise process $\phi_{LO}(t)$ is a Wiener process; therefore, the continuous-wave optical field of the LO laser has a Lorentzian spectrum with linewidth $\Delta \nu$ (cf. Fig.~\ref{system}b). The DSP consists of matched filtering (including EDC) modeled by $H^*(f)$, symbol-spaced sampling at the rate $B$, and the CPR algorithm. The sampled received symbol at CPR input (output) is denoted by $y_k$ ($\hat{y_k}$).\\ 

We denote the EEPN term at CPR input by $w_k$. It is \emph{defined} by the following equation
\begin{equation}\label{wk_before_CPR}
y_k = \sqrt{P} x_k e^{j\phi_k} + n_k + w_k,
\end{equation}
where $\phi_k=\phi_{LO}(kT_s)$ is the sampled LO phase, $n_k=n(kT_s)$ the sampled ASE noise, and $P=\sqrt{2P_lP_{in}}$ the received signal power after coherent mixer.

At the receiver, the CPR gives an estimate $\hat{\phi}_k$ of the carrier phase and rotates the received signal by this estimated phase $\hat{\phi}_k$ such that the signal after CPR is $\hat{y_k}=y_ke^{-j\hat{\phi}_k}$. The total distortion, including ASE, residual phase noise and residual EEPN after CPR, is denoted by $N_k$ and is written as
\begin{equation}
N_k = \hat{y}_k - \sqrt{P}x_k e^{j\phi_k}.
\end{equation}

Finally, we compute the variance $\sigma_N^2$ of the total noise $N_k$ and estimate the signal-to-noise ratio (SNR) of the received signal after CPR as $SNR = P/\sigma_N^2$.

\section{DSP-Aware EEPN characterization}\label{sec:DSPaware}

\subsection{EEPN identification before Carrier Phase Recovery}\label{details_stats}

We performed Monte Carlo simulations and studied statistical properties of the EEPN. We considered a polarized 64QAM channel transmission over EX3000 fiber (dispersion coefficient $D$~=~20.6~ps/nm/km) with central frequency $f_0$~=~194~THz and neglected nonlinear impairments. We set $P_{in}$~=~0~dBm, $P_l$~=~10~dBm, and $OSNR$~=~20~dB/0.1nm. We generated 10 sample functions of the LO phase noise process over a time window including $2^{17}$ symbols, and threw out 15000 symbols from each side of every simulated waveform, to avoid biasing the statistics by the numerical transients. In the simulations, we generated and saved the $x_k$, $\phi_k$ and $n_k$, such that once $y_k$ was produced, $w_k$ could be extracted without ambiguity using its definition in eq.~\eqref{wk_before_CPR}.

\begin{figure}[h!]
\centering
% This file was created by matplotlib2tikz v0.6.18.
\begin{tikzpicture}
\tikzstyle{every node}=[font=\scriptsize]
\begin{axis}[
axis on top,
tick pos=both,
xmajorgrids,
xmin=1000, xmax=1000000,
xminorgrids,
xmode=log,
xtick={10,100,1000,10000,100000,1000000,10000000},
xticklabels={,,$10^{3}$,$10^{4}$,$10^{5}$,$10^{6}$,},
ylabel={Noise variance},
ymajorgrids,
ymin=1e-09, ymax=1e-05,
yminorgrids,
ymode=log,
ytick={1e-10,1e-09,1e-08,1e-07,1e-06,1e-05,0.0001},
yticklabels={,$10^{-9}$,$10^{-8}$,$10^{-7}$,$10^{-6}$,$10^{-5}$,},
xlabel={LO linewidth $\Delta\nu$ [Hz]},
ylabel near ticks,
xlabel near ticks,
width=\linewidth,
height=6cm
]
\addplot [black, forget plot]
table [row sep=\\]{%
1000	1.6761e-09 \\
10000	1.6761e-08 \\
20000	3.3522e-08 \\
50000	8.3805e-08 \\
100000	1.6761e-07 \\
200000	3.3522e-07 \\
500000	8.3805e-07 \\
1000000	1.6761e-06 \\
};\label{Shieh49}
\addplot [black, dashed, forget plot]
table [row sep=\\]{%
1000	3.3522e-09 \\
10000	3.3522e-08 \\
20000	6.7044e-08 \\
50000	1.6761e-07 \\
100000	3.3522e-07 \\
200000	6.7044e-07 \\
500000	1.6761e-06 \\
1000000	3.3522e-06 \\
};\label{Shieh98}
\addplot [black, mark=*, mark size=4, mark options={solid}, only marks, forget plot]
table [row sep=\\]{%
1000	1.64577e-09 \\
10000	1.645344e-08 \\
20000	3.289731e-08 \\
50000	8.217253e-08 \\
100000	1.641104e-07 \\
200000	3.272967e-07 \\
500000	8.114473e-07 \\
1000000	1.600862e-06 \\
};\label{wk49}
\addplot [black, mark=*, mark size=4, mark options={solid,fill=white}, only marks, forget plot]
table [row sep=\\]{%
1000	3.234577e-09 \\
10000	3.233017e-08 \\
20000	6.462587e-08 \\
50000	1.6130598e-07 \\
100000	3.217555e-07 \\
200000	6.40102e-07 \\
500000	1.5751077e-06 \\
1000000	3.06605e-06 \\
};\label{wk98}
\path [draw=black, fill opacity=0] (axis cs:1000,1)
--(axis cs:1000000,1);

\path [draw=black, fill opacity=0] (axis cs:0,1e-09)
--(axis cs:0,1e-05);

\path [draw=black, fill opacity=0] (axis cs:1000,0)
--(axis cs:1000000,0);

\path [draw=black, fill opacity=0] (axis cs:1,1e-09)
--(axis cs:1,1e-05);

\coordinate (leg1) at (axis cs: 1100,9e-6);
\coordinate (leg2) at (axis cs: 7e4,4e-7);

\end{axis}

\draw[fill=white] (leg1) rectangle node[black]{
\begin{tabular}{clcl}
\multicolumn{4}{l}{Analytical EEPN variance $P\sigma_{EEPN}^2$ \cite{shieh2008}} \\
\ref{Shieh49} & 49 GBd & \ref{Shieh98} & 98 GBd \\
\multicolumn{4}{l}{Simulated EEPN term variance $\sigma_w^2$} \\
\ref{wk49} & 49 GBd & \ref{wk98} & 98 GBd \\
\end{tabular}
}(leg2);

\end{tikzpicture}
\caption{Analytical and numerical EEPN variance evolution as a function of LO linewidth, for 49 and 98~GBd 64QAM transmission through 6600~km EX3000}
\label{varwk}
\end{figure}
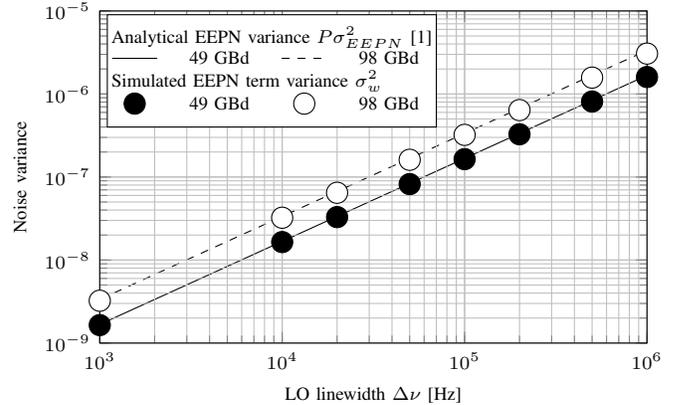

For a link of $L$~=~6600~km, Fig.~\ref{varwk} compares the analytical variance computed in \cite{shieh2008} for a baudrate $B$ of 49~GBd (98~GBd) drawn in solid (dashed) line, and the variance $\sigma_w^2$ of the EEPN term $w_k$ computed from the numerical simulations, for 49~GBd (black circles) and 98~GBd (white circles), as functions of the LO linewidth $\Delta\nu$. We show that, before CPR, our simulations and identification of EEPN term as $w_k$ match the analytical model of \cite{shieh2008} for the EEPN variance 
\begin{equation}\label{varShieh}
\sigma_{EEPN}^2 = \frac{\pi c D L B \Delta\nu}{2f_0^2}.
\end{equation}

Note that \eqref{varShieh} is valid for normalized transmitted symbols and therefore, $\sigma^2_{EEPN}$ has been here multiplied by the signal power $P$ to meet our definition of EEPN term, which does not use power normalization.

Fig.~\ref{pyPDF}a shows the normalized probability distribution function (PDF) of the real part of the EEPN term $w_k$ at CPR input (dashed line), as well as the Gaussian fit (solid line), showing that $w_k$ is not Gaussian. The normalized PDF of the real part of the total noise $n_k+w_k$ is plotted in fig.~\ref{pyPDF}b (dashed line), and fits a Gaussian shape (solid line). Even if the EEPN part can not be considered as a Gaussian noise, the total noise accounting for ASE and EEPN remains Gaussian in the considered setup. 

\begin{figure}[h!]
\begin{minipage}{0.49\linewidth}
\begin{center}
\input{pyPDF}
\scriptsize (a)
\end{center}
\end{minipage}\hfill
\begin{minipage}{0.49\linewidth}
\begin{center}
\input{pyPDF2}
\scriptsize (b)
\end{center}
\end{minipage}
\caption{Probability density function (PDF) of the real part of EEPN distortions at carrier phase recovery (CPR) input $w_k$ (a) and total distortions at carrier phase recovery (CPR) input $n_k+w_k$ (b). Numerical simulations were performed for a 49~GBd 64QAM transmission through 6600~km EX3000, neglecting nonlinearity, with OSNR equal to 20~dB/0.1nm and LO linewidth set to 200~kHz}
\label{pyPDF}
\end{figure}

For 49~GBd transmission over 6600 or 13419~km, fig.~\ref{pyXcorr} shows the normalized amplitude of the cross-correlation $R_n$ between the EEPN distortions with ideal data remodulation $w_kx_k^*$ and the LO field $e^{j\phi_k}$, defined as 
\begin{equation}
R_n = \langle (w_k x_k^*)^* e^{j\phi_{k+n}} \rangle = \langle w_k^* x_k e^{j\phi_{k+n}}\rangle
\end{equation}

We observe a one-sided width at half maximum of around 400 symbols for 6600~km, and 760 symbols for 13419~km. This shows that the EEPN is a colored noise which depends on link properties and is correlated to the LO phase noise. Therefore, taking into account the influence of the phase recovery DSP in the analysis is a must.

\begin{figure}[h!]
\begin{minipage}{0.49\linewidth}
\begin{center}
% This file was created by matplotlib2tikz v0.6.18.
\begin{tikzpicture}
\tikzstyle{every node}=[font=\figuretextsize]
\begin{axis}[
axis on top,
tick pos=both,
xlabel={Symbols},
xmajorgrids,
xmin=0, xmax=4000,
ylabel={$|R_n|/|R_0|$},
ymajorgrids,
ymin=0, ymax=1,
ytick={0,0.25,0.5,0.75,1},
yticklabels={0,,,,1},
xtick={0,1000,2000,3000,4000},
xticklabels={0,1000,2000,3000,4000},
y label style={at={(axis description cs:.35,.5)},anchor=south},
width=.8\textwidth,
height=3cm,
scale only axis
]
\addplot [black, forget plot]
table [row sep=\\]{%
0	1 \\
40	0.940056675992503 \\
80	0.882843279723698 \\
120	0.827378069328501 \\
160	0.77398180039057 \\
200	0.722591382530616 \\
240	0.67316424098115 \\
280	0.625816359739731 \\
320	0.579856015177552 \\
360	0.536224357882946 \\
400	0.493910716090417 \\
440	0.453318566613918 \\
480	0.41419949928719 \\
520	0.376693955865973 \\
560	0.341343556054521 \\
600	0.30870042326991 \\
640	0.278023078946224 \\
680	0.248878283567945 \\
720	0.221398733738377 \\
760	0.196722975281319 \\
800	0.174370139313229 \\
840	0.154253169581634 \\
880	0.137421529058294 \\
920	0.123351791609426 \\
960	0.111899454484007 \\
1000	0.10307745052794 \\
1040	0.0967442960147011 \\
1080	0.0915145741063505 \\
1120	0.0878438522801949 \\
1160	0.0862353122790449 \\
1200	0.084924614908689 \\
1240	0.0839155931035863 \\
1280	0.0841586811038552 \\
1320	0.0838947293112567 \\
1360	0.0842844894475177 \\
1400	0.0847354340452394 \\
1440	0.0848019511382136 \\
1480	0.0847246742698551 \\
1520	0.0849444649282736 \\
1560	0.0855861094198625 \\
1600	0.0861395413902557 \\
1640	0.0867921324280068 \\
1680	0.0870675669462001 \\
1720	0.0874849594167662 \\
1760	0.0883991251123159 \\
1800	0.088797435068556 \\
1840	0.0895990857491867 \\
1880	0.0903385097523834 \\
1920	0.0908022831676666 \\
1960	0.091287884260903 \\
2000	0.0921146188339143 \\
2040	0.0932216920092013 \\
2080	0.0940141673107779 \\
2120	0.0944284724828015 \\
2160	0.0954566571391707 \\
2200	0.0955479269114765 \\
2240	0.0955221053245089 \\
2280	0.0953750898052918 \\
2320	0.094723842416522 \\
2360	0.0938580895987379 \\
2400	0.0934303754169212 \\
2440	0.0930638113713009 \\
2480	0.0925217896386347 \\
2520	0.0919705783726912 \\
2560	0.0900860331403982 \\
2600	0.0894972683018901 \\
2640	0.0888654043307474 \\
2680	0.0881915825975143 \\
2720	0.0872906191430332 \\
2760	0.0868730487162711 \\
2800	0.0858291693897889 \\
2840	0.0849607072345297 \\
2880	0.0845027090207881 \\
2920	0.0845977495510593 \\
2960	0.084611193725945 \\
3000	0.0846116094810373 \\
3040	0.0847751809943893 \\
3080	0.0847510799731989 \\
3120	0.0846682262928452 \\
3160	0.0846725437653388 \\
3200	0.0850304138113683 \\
3240	0.0861192565935558 \\
3280	0.0871458066104667 \\
3320	0.0872566025864506 \\
3360	0.0875524389398191 \\
3400	0.0872949035248846 \\
3440	0.0870590236861703 \\
3480	0.0864200132502873 \\
3520	0.0852418540925787 \\
3560	0.0850394214737248 \\
3600	0.0842704486633113 \\
3640	0.0837398190134208 \\
3680	0.0829985311869279 \\
3720	0.0825138952497784 \\
3760	0.0824554224107636 \\
3800	0.0824848901471679 \\
3840	0.0824829660381132 \\
3880	0.0824144004970918 \\
3920	0.082244769761661 \\
3960	0.0818043295153673 \\
};
\path [draw=black, fill opacity=0] (axis cs:0,1)
--(axis cs:4000,1);

\path [draw=black, fill opacity=0] (axis cs:0,0)
--(axis cs:0,1);

\path [draw=black, fill opacity=0] (axis cs:1,0)
--(axis cs:1,1);

\path [draw=black, fill opacity=0] (axis cs:0,0)
--(axis cs:4000,0);

\coordinate (a) at (axis cs:2100,0.9);

\coordinate (b0) at (axis cs: 0,0.5);
\coordinate (b1) at (axis cs: 420,0.5);
\coordinate (b2) at (axis cs: 420,0);
\end{axis}

\draw (a) node{$R_n = \langle w_k^*x_k e^{j\phi_{k+n}}\rangle$};
\draw[dashed] (b0) -- (b1) node[right]{\begin{tabular}{l}$L$=6600~km \\One-sided width \\ at half maximum \\ around 400 symbols\end{tabular}} -- (b2);
\end{tikzpicture}
\scriptsize (a)
\end{center}
\end{minipage}\hfill
\begin{minipage}{0.49\linewidth}
\begin{center}
% This file was created by matplotlib2tikz v0.6.18.
\begin{tikzpicture}
\tikzstyle{every node}=[font=\figuretextsize]
\begin{axis}[
axis on top,
tick pos=both,
xlabel={Symbols},
xmajorgrids,
xmin=0, xmax=4000,
ylabel={$|R_n|/|R_0|$},
ymajorgrids,
ymin=0, ymax=1,
ytick={0,0.25,0.5,0.75,1},
yticklabels={0,,,,1},
xtick={0,1000,2000,3000,4000},
xticklabels={0,1000,2000,3000,4000},
y label style={at={(axis description cs:.35,.5)},anchor=south},
width=.8\textwidth,
height=3cm,
scale only axis
]
\addplot [black, forget plot]
table [row sep=\\]{%
0	1 \\
40	0.970020101381756 \\
80	0.940672576222281 \\
120	0.911763025721787 \\
160	0.883509561173755 \\
200	0.8553853857121 \\
240	0.827708599672349 \\
280	0.80017351157187 \\
320	0.773091390894614 \\
360	0.746374739823923 \\
400	0.720376941249687 \\
440	0.694751027750075 \\
480	0.670012497354297 \\
520	0.645728859053367 \\
560	0.621921832397144 \\
600	0.598044511981215 \\
640	0.574738365989503 \\
680	0.552000135620454 \\
720	0.529468281453943 \\
760	0.507736283447893 \\
800	0.486094383657637 \\
840	0.465168659195968 \\
880	0.444891218726179 \\
920	0.42478382544627 \\
960	0.404820609335455 \\
1000	0.385740259288283 \\
1040	0.367344919059225 \\
1080	0.349677701380992 \\
1120	0.33244708595055 \\
1160	0.315678415096864 \\
1200	0.299374381969408 \\
1240	0.283649045730606 \\
1280	0.268574032643887 \\
1320	0.254253054590134 \\
1360	0.240540174139852 \\
1400	0.227474082102226 \\
1440	0.21518163789964 \\
1480	0.203273802300053 \\
1520	0.192224993306384 \\
1560	0.182254055754704 \\
1600	0.172694542218434 \\
1640	0.164020950645807 \\
1680	0.15578257797151 \\
1720	0.148012001633467 \\
1760	0.140992374076673 \\
1800	0.134362063135582 \\
1840	0.128899680174903 \\
1880	0.123756038972021 \\
1920	0.119867265038609 \\
1960	0.116701212661607 \\
2000	0.11448065321135 \\
2040	0.11297525673296 \\
2080	0.111772695010251 \\
2120	0.110920856228464 \\
2160	0.110712865563649 \\
2200	0.110366677589518 \\
2240	0.110438822324371 \\
2280	0.110578813526496 \\
2320	0.110529826143981 \\
2360	0.110288417347462 \\
2400	0.110622766197631 \\
2440	0.110986096136293 \\
2480	0.111499930249245 \\
2520	0.112211565493418 \\
2560	0.11235034818722 \\
2600	0.112339723463076 \\
2640	0.112875922883786 \\
2680	0.113370603240981 \\
2720	0.113694682362408 \\
2760	0.114107018111212 \\
2800	0.114210329912195 \\
2840	0.114059707002667 \\
2880	0.114122887332402 \\
2920	0.114287560631909 \\
2960	0.114237378802465 \\
3000	0.114183038540349 \\
3040	0.114128360249703 \\
3080	0.114026137342669 \\
3120	0.113542498677124 \\
3160	0.11328877481057 \\
3200	0.113165256033217 \\
3240	0.113142906759107 \\
3280	0.112981651380657 \\
3320	0.112916661300027 \\
3360	0.11309346727091 \\
3400	0.112750873959732 \\
3440	0.112320810577203 \\
3480	0.112134163304011 \\
3520	0.111823036857171 \\
3560	0.11154473307398 \\
3600	0.110980126398117 \\
3640	0.110712930135673 \\
3680	0.110332266237834 \\
3720	0.10988217457323 \\
3760	0.109500961786504 \\
3800	0.109100383638339 \\
3840	0.108919815655241 \\
3880	0.108583605064303 \\
3920	0.108100328908692 \\
3960	0.108097730714559 \\
};

\path [draw=black, fill opacity=0] (axis cs:0,1)
--(axis cs:4000,1);

\path [draw=black, fill opacity=0] (axis cs:0,0)
--(axis cs:0,1);

\path [draw=black, fill opacity=0] (axis cs:1,0)
--(axis cs:1,1);

\path [draw=black, fill opacity=0] (axis cs:0,0)
--(axis cs:4000,0);

\coordinate (a) at (axis cs:2100,0.9);

\coordinate (b0) at (axis cs: 0,0.5);
\coordinate (b1) at (axis cs: 760,0.5);
\coordinate (b2) at (axis cs: 760,0);
\end{axis}

\draw (a) node{$R_n = \langle w_k^*x_k e^{j\phi_{k+n}}\rangle$};
\draw[dashed] (b0) -- (b1) node[right]{\begin{tabular}{l}$L$=13419~km\\One-sided width \\ at half maximum \\ around 760 symbols\end{tabular}} -- (b2);
\end{tikzpicture}
\scriptsize (b)
\end{center}
\end{minipage}
\caption{Signal-EEPN cross-correlation for a 49~GBd 64QAM transmission through (a) 6600~km or (b) 13419~km EX3000, neglecting nonlinearity, with OSNR equal to 20~dB/0.1nm and LO linewidth set to 200~kHz}
\label{pyXcorr}
\end{figure}

\subsection{Carrier Phase Recovery influence}\label{details_CPR}

To study the behavior of CPR algorithms in the presence of EEPN, we consider three different cases of CPR implementation to describe the performance of the system. First, we perform a LO phase cancellation, i.e. we assume $\hat{\phi}_k=\phi_k$, which requires a perfect knowledge of the random process $\phi_k$. Next, we use ideal data remodulation (IDR), namely, we assume perfect knowledge of the transmitted symbols $x_k$ and compute the estimated phase over an averaging block of length $N=2l+1$ as $\hat{\phi}_k = \angle \sum_{k=-l}^l y_k x_k^*$. Finally, to study the performance of a realistic transmission use-case, in the third case the estimated phase $\hat{\phi}_k$ is computed as the output of the BPS algorithm with an averaging block of length $N=2l+1$. Note that when using BPS, in this paper, all cycle slips are removed to avoid performance degradation from ambiguity coming from constellation symmetry.

\begin{figure}
\centering
% This file was created by matplotlib2tikz v0.6.18.
\begin{tikzpicture}
\tikzstyle{every node}=[font=\figuretextsize]
\begin{axis}[
axis on top,
tick pos=both,
xlabel={CPR averaging length [symbols]},
xmajorgrids,
xmin=10, xmax=10000,
xminorgrids,
xmode=log,
ylabel={SNR [dB]},
ylabel near ticks,
xlabel near ticks,
ymajorgrids,
ymin=11.5, ymax=14.5,
ytick={12,12.5,13,13.5,14},
yticklabels={12,,13,,14},
yminorgrids,
width=\linewidth,
height=6cm
]
\addplot [ultra thick, black, forget plot]
table [row sep=\\]{%
10	14.0668 \\
10000	14.0668 \\
};\label{SNR_CPE_D0}
\addplot [dotted, black, mark=diamond*, mark size=4, mark options={solid,fill=white}, forget plot]
table [row sep=\\]{%
30	14.0659218145321 \\
40	14.0659218145321 \\
50	14.0659218145321 \\
70	14.0659218145321 \\
100	14.0659218145321 \\
200	14.0659218145321 \\
400	14.0659218145321 \\
700	14.0659218145321 \\
1000	14.0659218145321 \\
1200	14.0659218145321 \\
1600	14.0659218145321 \\
2000	14.0659218145321 \\
2500	14.0659218145321 \\
3000	14.0659218145321 \\
5000	14.0659218145321 \\
};
\label{SNR_CPE_D0_PT}

\addplot [dotted, black, dotted, mark=diamond*, mark size=4, mark options={solid}, forget plot]
table [row sep=\\]{%
30	14.0890320195689 \\
40	14.0758015627676 \\
50	14.0658512517419 \\
70	14.0501213081879 \\
100	14.0317313808651 \\
200	13.9786226986363 \\
400	13.8834193377148 \\
700	13.7484626636496 \\
1000	13.6161227632432 \\
1200	13.529595689375 \\
1600	13.3627222813604 \\
2000	13.1920104513912 \\
2500	12.9890268082019 \\
3000	12.8028943041311 \\
5000	12.2060777300667 \\
};
\label{SNR_CPE_D0_IDR}

\addplot [dotted, lightgray!66.928104575163388!black, mark=diamond*, mark size=4, mark options={solid,draw=black}, forget plot]
table [row sep=\\]{%
30	12.7096231853822 \\
40	13.288920393646 \\
50	13.4929619231046 \\
70	13.6345349686285 \\
100	13.7228586764622 \\
200	13.7994528170591 \\
400	13.7731151400373 \\
700	13.6828854361614 \\
1000	13.5758340293024 \\
1200	13.5053972097443 \\
1600	13.3609938040096 \\
2000	13.2118483563294 \\
2500	13.0406783275455 \\
3000	12.8825596180541 \\
5000	12.3583122806926 \\
};
\label{SNR_CPE_D0_BPS}

\addplot [ultra thick, black, dashed, forget plot]
table [row sep=\\]{%
10	12.5209 \\
10000	12.5209 \\
};
\label{SNR_CPE_ASEeq}

\addplot [dotted, black, mark=*, mark size=4, mark options={solid,fill=white}, forget plot]
table [row sep=\\]{%
30	12.5298201796414 \\
40	12.5298201796414 \\
50	12.5298201796414 \\
70	12.5298201796414 \\
100	12.5298201796414 \\
200	12.5298201796414 \\
400	12.5298201796414 \\
700	12.5298201796414 \\
1000	12.5298201796414 \\
1200	12.5298201796414 \\
1600	12.5298201796414 \\
2000	12.5298201796414 \\
2500	12.5298201796414 \\
3000	12.5298201796414 \\
5000	12.5298201796414 \\
};
\label{SNR_CPE_D1_PT}

\addplot [black, dotted, mark=*, mark size=4, mark options={solid}, forget plot]
table [row sep=\\]{%
30	13.0299039002227 \\
40	13.0192867740136 \\
50	13.0121372757695 \\
70	13.0039721423636 \\
100	12.9978174218415 \\
200	12.9901188478291 \\
400	12.9865560425498 \\
700	12.9805413753766 \\
1000	12.9693872823628 \\
1200	12.9575519463601 \\
1600	12.9180979061977 \\
2000	12.8554304483221 \\
2500	12.7481266724221 \\
3000	12.6208655613954 \\
5000	12.1221026217505 \\
};
\label{SNR_CPE_D1_IDR}

\addplot [dotted, lightgray!66.928104575163388!black, mark=*, mark size=4, mark options={solid,draw=black}, forget plot]
table [row sep=\\]{%
20	9.88684945556992 \\
30	11.6509438586856 \\
40	12.1765661057921 \\
50	12.408649180396 \\
70	12.5766926842136 \\
100	12.6831984097313 \\
200	12.8092369001288 \\
400	12.8825083716139 \\
700	12.9090367017965 \\
1000	12.9133335078849 \\
1200	12.9098005223229 \\
1600	12.885976522627 \\
2000	12.8413323775893 \\
2500	12.760757667269 \\
3000	12.6618179487274 \\
5000	12.2614109600746 \\
};
\label{SNR_CPE_D1_BPS}

\coordinate (a) at (axis cs: 8000,14);
\coordinate (b) at (axis cs: 8000,12.9);
\coordinate (bb) at (axis cs: 1000,12.9);
\coordinate (c) at (axis cs: 8000,12.89);
\coordinate (d) at (axis cs: 8000,12.55);
\coordinate (leg1) at (axis cs: 72,12.05);
\coordinate (leg2) at (axis cs: 9200,11.55);
\end{axis}

\draw[ thick, black, dashed] (b) -- (bb);
\draw[->,>=stealth,ultra thick, black] (a) -- (b) node[midway,left]{(1)};
\draw[->,>=stealth,ultra thick, black] (c) -- (d) node[midway,left]{(2)};

\draw[fill=white] (leg1) rectangle node[black]{
\begin{tabular}{ll}
(1)&EEPN penalty\\
(2)&Analytical overestimation
\end{tabular}
}(leg2);

\end{tikzpicture}

\centering
\hrule
{\figuretextsize
\begin{tabular}{lll}
\multicolumn{3}{l}{\bf Analytical:} \\
\parbox{1cm}{\centering \ref{SNR_CPE_D0}} ASE only & \multicolumn{2}{l}{\ref{SNR_CPE_ASEeq} ASE + equivalent AWGN \cite{shieh2008}} \\
\multicolumn{3}{l}{\bf Simulation with ASE, without EEPN (dispersion kept to 0):} \\
\parbox{1cm}{\centering \ref{SNR_CPE_D0_PT}} LO phase cancellation & \ref{SNR_CPE_D0_IDR} IDR & \ref{SNR_CPE_D0_BPS} BPS \\
\multicolumn{3}{l}{\bf Simulation with ASE and EEPN:} \\
\parbox{1cm}{\centering \ref{SNR_CPE_D1_PT}} LO phase cancellation & \ref{SNR_CPE_D1_IDR} IDR & \ref{SNR_CPE_D1_BPS} BPS \\
\end{tabular}
}
\hrule
\caption{SNR vs. CPR averaging length, for LO phase cancellation, IDR and BPS carrier phase estimation of a 49~GBd 64QAM transmission through 6600~km EX3000, neglecting nonlinearity, with OSNR equal to 20~dB/0.1nm and LO linewidth set to 200~kHz}
\label{SNR_CPE_64QAM}
\end{figure}

Assuming the same setup as in fig.~\ref{pyPDF} and with the three different CPR techniques, we computed the received signal $\hat{y}_k$ after CPR and we plot the estimated SNR in Fig.~\ref{SNR_CPE_64QAM}. The solid black line is the theoretical SNR for a fixed OSNR of 20~dB/0.1nm, assuming that ASE is the only source of noise. Next, we turn on LO phase noise but keep dispersion to zero: just as a numerical sanity check, white diamonds corresponding to LO phase cancellation (when $\phi_k$ is used as $\hat{\phi}_k$) meet the solid line of the theoretical SNR; black diamonds correspond to ideal data remodulation (IDR) whereas gray diamonds show the result of the BPS algorithm over a block of $N=2l+1$ symbols. For short CPR averaging, the black diamonds reach the theoretical performance, showing that LO phase noise is completely removed by data remodulation. As the block length increases, the LO phase noise tracking is degraded by the averaging process and the performance decreases. The gray diamonds corresponding to BPS exhibit an optimal averaging length around 200 symbols. Below this optimum, phase estimation is degraded by ASE noise which is not sufficiently averaged by the CPR process. Beyond this optimum, the averaging window is too long and does not allow to follow the LO phase noise, hence the BPS performance curve converges to that of the IDR curve, and both IDR and BPS performance degrade due to not following the phase noise Wiener process.

Next, we turn on the dispersion (hence EEPN): white circles represent the performance with CPR by LO phase cancellation, black circles correspond to the IDR with perfect knowledge of transmitted symbols, and gray circles are the result of BPS. We observe that the phase recovery by LO phase cancellation meets the case where EEPN is modeled as an equivalent AWGN (dashed line) with the variance computed in \cite{shieh2008} and is \emph{not} the best strategy to perform CPR in presence of EEPN. Even if we have perfect knowledge of the LO phase noise, as shown in \cite{farhoudi2012}, using LO phase noise as phase estimate $\hat{\phi}_k$ is non-optimal as it ignores that the phase to be removed by the CPR is a combination of LO phase noise $\phi_k$, transmitted symbols $x_k$ and CDC equalizer transfer function $H(f)$. Consequently, the black circles representing ideal data remodulation exhibit the best performance among the three considered cases in the presence of EEPN: for short averaging length, it reaches a SNR level which is 0.6~dB greater than that achieved by AWGN equivalent noise model; the performance decreases slowly when the averaging length increases (only 0.2~dB when the block size varies from 30 to 1000), before decreasing strongly when the averaging process does not allow to track the phase variations anymore. Finally, we show that the BPS algorithm (gray circles) exhibits an optimal performance which almost reaches the ideal data remodulation case, outperforming the analytical prediction by 0.4~dB when the averaging block length is set to 1000. In the presence of EEPN, the optimal performance is attained for longer averaging block length compared to phase-noise-only case. This change of behavior of the BPS algorithm in the presence of EEPN can be considered as a signature of EEPN non-white nature, showing that the BPS must be taken into account in the study of EEPN impact. The SNR offset labeled as (1) on Fig. 5, shows the true EEPN penalty with respect to optimized BPS. This penalty could be avoided if an ideal EEPN compensator were available. The SNR offset labeled as (2) is the the amount by which the closed form formula, as per (3) overestimates the EEPN penalty.

\subsection{Impact of modulation format choice}\label{details_MF}

\begin{figure}
\centering
% This file was created by matplotlib2tikz v0.6.18.
\begin{tikzpicture}
\tikzstyle{every node}=[font=\figuretextsize]
\begin{axis}[
axis on top,
tick pos=both,
xlabel={CPR averaging length [symbols]},
xmajorgrids,
xmin=10, xmax=10000,
xminorgrids,
xmode=log,
ylabel={SNR [dB]},
ylabel near ticks,
xlabel near ticks,
ymajorgrids,
ymin=11, ymax=14,
yminorgrids,
width=\linewidth,
height=5cm
]
\addplot [dotted, black, mark=*, mark size=4, mark options={solid,fill=black}, forget plot]
table [row sep=\\]{%
30	13.010940162204 \\
40	12.9999492292058 \\
50	12.9938585941263 \\
70	12.9853372978095 \\
100	12.9785946782826 \\
200	12.9726496475655 \\
400	12.9679476227227 \\
700	12.9624865328263 \\
1000	12.951184463736 \\
1200	12.9394477452571 \\
1600	12.900232026435 \\
2000	12.8398003920125 \\
2500	12.7324137725055 \\
3000	12.6153699297586 \\
5000	12.1290864338211 \\
};\label{SNR_CPE_MF_QPSK}
%\addplot [dotted, black, mark=*, mark size=4, mark options={solid,fill=white}, forget plot]
%table [row sep=\\]{%
%20	12.8167742287808 \\
%30	12.9299453465093 \\
%40	12.9494765161214 \\
%50	12.9521020705272 \\
%70	12.9550510259608 \\
%100	12.9567838217197 \\
%200	12.9569568399719 \\
%400	12.9566110553975 \\
%700	12.952453776171 \\
%1000	12.9429664754102 \\
%1200	12.9333598461724 \\
%1600	12.8979108530722 \\
%2000	12.8456599938017 \\
%2500	12.7509931327557 \\
%3000	12.654998731472 \\
%5000	12.1581072637951 \\
%};\label{SNR_CPE_MF_16QAM}
\addplot [dotted, lightgray!66.928104575163388!black, mark=*, mark size=4, mark options={solid,draw=black}, forget plot]
table [row sep=\\]{%
30	11.6509438586856 \\
40	12.1765661057921 \\
50	12.408649180396 \\
70	12.5766926842136 \\
100	12.6831984097313 \\
200	12.8092369001288 \\
400	12.8825083716139 \\
700	12.9090367017965 \\
1000	12.9133335078849 \\
1200	12.9098005223229 \\
1600	12.885976522627 \\
2000	12.8413323775893 \\
2500	12.760757667269 \\
3000	12.6618179487274 \\
5000	12.2614109600746 \\
};\label{SNR_CPE_MF_64QAM}
\addplot [dotted, black, mark=diamond*, mark size=4, mark options={solid}, forget plot]
table [row sep=\\]{%
30	5.85370996701084 \\
40	7.21338060609691 \\
50	8.33368125636654 \\
70	9.88877647196254 \\
100	11.1955093774784 \\
200	12.4899893549732 \\
400	12.8320001224133 \\
700	12.9237095585423 \\
1000	12.9133959859934 \\
1200	12.8914499019266 \\
1600	12.8274096122816 \\
2000	12.7447734513418 \\
2500	12.612865653865 \\
3000	12.4607997731148 \\
5000	11.8683077642393 \\
};\label{SNR_CPE_MF_pcs64}

\addplot [ultra thick, black, dashed, forget plot]
table [row sep=\\]{%
10	12.5209 \\
10000	12.5209 \\
};
\label{SNR_CPE_MF_ASEeq}

\coordinate (leg1) at (axis cs:11,13.9);
\coordinate (leg2) at (axis cs: 3100,13.2);
\end{axis}

\draw[fill=white] (leg1) rectangle node[black]{
\begin{tabular}{l}
\ref{SNR_CPE_MF_QPSK} QPSK 
\ref{SNR_CPE_MF_64QAM} 64QAM  
\ref{SNR_CPE_MF_pcs64} PCS64QAM \\
\ref{SNR_CPE_MF_ASEeq} (Analytical) ASE + equivalent AWGN \cite{shieh2008} 
\end{tabular}
} (leg2);

\end{tikzpicture}
\caption{SNR vs. CPR averaging length, for optimized as BPS carrier phase estimation of a 49~GBd transmission using QPSK, 64QAM or PCS64QAM, through 6600~km EX3000, neglecting nonlinearity, with OSNR equal to 20~dB/0.1nm.}
\label{SNR_CPE_formats}
\end{figure}
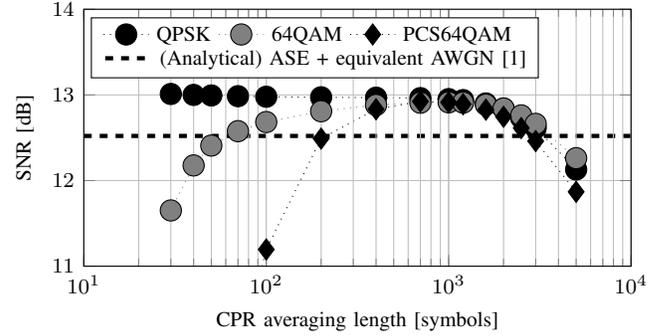

Previous analysis was performed assuming 64QAM modulation. Now we study the influence of modulation format on BPS performance in presence of EEPN. Assuming the same setup as the one resulted in the curves of Fig.\ref{SNR_CPE_64QAM}, we plot in Fig.\ref{SNR_CPE_formats}, in presence of EEPN, the SNR after BPS for QPSK (black circles), 64QAM (gray circles) and probabilistically shaped PCS64QAM, with source entropy equal to 5.4 bits/symbol (black diamonds). The theoretical performance with EEPN modeled as equivalent AWGN is drawn in dashed line. Compared to 64QAM, QPSK shows better resilience to CPR averaging length variation, with a stable SNR performance of almost 13~dB with moving average filter lengths varying from 50 to 1000. For the short averaging length where the phase estimation process is limited by the ASE noise, decision errors arise more often with constellations with high cardinality and small distances between symbols. For PCS64QAM, this degradation drastically falls at low averaging lengths, indicating that probabilistically shaped format imposes even tighter constraints on the CPR parameter than the high cardinality unshaped formats. However, the maximum achieved SNR for all the constellations is very similar, provided the moving average legnth is separately optimized four each format. The block averaging length must be optimized in each configuration and under this condition, in this example, the degradation from the advanced formats remains negligible at the optimal point of operation.

\begin{figure}
\centering
% This file was created by matplotlib2tikz v0.6.18.
\begin{tikzpicture}
\tikzstyle{every node}=[font=\figuretextsize]
\begin{axis}[
axis on top,
tick pos=both,
xlabel={Laser linewidth [Hz]},
xmajorgrids,
xmin=1000, xmax=1000000,
xminorgrids,
xmode=log,
ylabel={SNR [dB]},
ylabel near ticks,
xlabel near ticks,
ymajorgrids,
ymin=7, ymax=15,
yminorgrids,
height=5cm,
width=\linewidth
]
\addplot [black, dotted, mark=*, mark size=4, mark options={solid}, forget plot]
table [row sep=\\]{%
1000	14.0915523876313 \\
10000	14.0378375822793 \\
20000	13.9781800397958 \\
50000	13.8040404644591 \\
100000	13.5308729689039 \\
200000	13.0307645399947 \\
500000	11.8134114232966 \\
1000000	10.3035459082493 \\
}; \label{pySNR_LW_MF_qpsk}
%\addplot [black, dotted, mark=*, mark size=4, mark options={solid,fill=white}, forget plot]
%table [row sep=\\]{%
%1000	14.0604129903211 \\
%10000	14.003797381213 \\
%20000	13.9416277633367 \\
%50000	13.757922157854 \\
%100000	13.4730883259924 \\
%200000	12.9569568399719 \\
%500000	11.7142482890335 \\
%1000000	10.2131501085462 \\
%}; \label{pySNR_LW_MF_16QAM}
\addplot [white!50.196078431372548!black, dotted, mark=*, mark size=4, mark options={solid,draw=black}, forget plot]
table [row sep=\\]{%
1000	14.0299892097178 \\
10000	13.9653294457799 \\
20000	13.9026467022844 \\
50000	13.7165707663479 \\
100000	13.430528996752 \\
200000	12.9133335078849 \\
500000	11.6719787489992 \\
1000000	10.1818337175296 \\
}; \label{pySNR_LW_MF_64QAM}
\addplot [black, dotted, mark=diamond*, mark size=4, mark options={solid}, forget plot]
table [row sep=\\]{%
1000	14.039234760234 \\
10000	13.982446693325 \\
20000	13.9236072938707 \\
50000	13.7486025924775 \\
100000	13.4688501132657 \\
200000	12.9237095585423 \\
500000	11.4618214866786 \\
1000000	9.70045185085186 \\
}; \label{pySNR_LW_MF_PCS}
\addplot [very thick, black, dashed, forget plot]
table [row sep=\\]{%
1000	14.0691 \\
10000	13.9863 \\
20000	13.8961 \\
50000	13.6363 \\
100000	13.235 \\
200000	12.529 \\
500000	10.9151 \\
1000000	9.1044 \\
2000000	6.8464 \\
5000000	3.3902 \\
}; \label{pySNR_LW_MF_shieh}
\addplot [very thick, black, forget plot]
table [row sep=\\]{%
1000	14.0784 \\
5000000	14.0784 \\
}; \label{pySNR_LW_MF_theory}

\addplot [black, dotted, mark=*, mark size=4, mark options={solid}, forget plot]
table [row sep=\\]{%
1000	11.0935835531456 \\
10000	11.0434088456843 \\
20000	10.9875537076018 \\
50000	10.8254551513113 \\
100000	10.5687835742109 \\
200000	10.0993376194625 \\
500000	8.93909438755989 \\
1000000	7.49475175482444 \\
};
%\addplot [black, dotted, mark=*, mark size=4, mark options={solid,fill=white}, forget plot]
%table [row sep=\\]{%
%1000	11.0383946887329 \\
%10000	10.9893992190535 \\
%20000	10.9346114676441 \\
%50000	10.7720561884543 \\
%100000	10.5143010086867 \\
%200000	10.0426136865611 \\
%500000	8.88544379006145 \\
%1000000	7.47670437848046 \\
%};
\addplot [white!50.196078431372548!black, dotted, mark=*, mark size=4, mark options={solid,draw=black}, forget plot]
table [row sep=\\]{%
1000	11.0310440688853 \\
10000	10.9817530328279 \\
20000	10.9281025627095 \\
50000	10.7642885544045 \\
100000	10.5071237022361 \\
200000	10.0364529818827 \\
500000	8.8887034080395 \\
1000000	7.49135867319809 \\
};
\addplot [black, dotted, mark=diamond*, mark size=4, mark options={solid}, forget plot]
table [row sep=\\]{%
1000	11.0193178358031 \\
10000	10.9648541696645 \\
20000	10.9043065984304 \\
50000	10.7223963275806 \\
100000	10.4339195936345 \\
200000	9.90482409925812 \\
500000	8.65763951511255 \\
1000000	7.18756077877815 \\
};
\addplot [very thick, black, dashed, forget plot]
table [row sep=\\]{%
1000	11.0588 \\
10000	10.976 \\
20000	10.8858 \\
50000	10.626 \\
100000	10.2247 \\
200000	9.5187 \\
500000	7.9048 \\
1000000	6.0941 \\
2000000	3.8361 \\
5000000	0.37988 \\
};
\addplot [very thick, black, forget plot]
table [row sep=\\]{%
1000	11.0681 \\
5000000	11.0681 \\
};
\coordinate (leg1) at (axis cs:1100,9.1);
\coordinate (leg2) at (axis cs: 210000,7.1);
\coordinate (a) at (axis cs: 2500, 14);
\coordinate (b) at (axis cs: 3300, 11);
\end{axis}

\draw (a) ellipse  (0.1cm and 0.4cm) node[below right]{49 GBd};
\draw (b) ellipse  (0.1cm and 0.4cm) node[above right]{98 GBd};

\end{tikzpicture}

\centering
\hrule
{\figuretextsize
\begin{tabular}{clclcl}
\multicolumn{6}{l}{\bf Analytical:} \\
\ref{pySNR_LW_MF_theory} & ASE only & \ref{pySNR_LW_MF_shieh} & \multicolumn{3}{l}{ASE + equivalent AWGN \cite{shieh2008}}\\
\multicolumn{6}{l}{\bf Simulation with ASE and EEPN:} \\
\ref{pySNR_LW_MF_qpsk} & QPSK & \ref{pySNR_LW_MF_64QAM} & 64QAM & \ref{pySNR_LW_MF_PCS} & PCS64QAM 
\end{tabular}
}
\hrule
\caption{SNR vs. LO laser linewidth, for optimized BPS as CPR of a 49~GBd or 98~GBd transmission using QPSK, 64QAM or PCS64QAM, through 6600~km EX3000, neglecting nonlinearity, with OSNR equal to 20~dB/0.1nm.}
\label{SNR_LW_MF}
\end{figure}
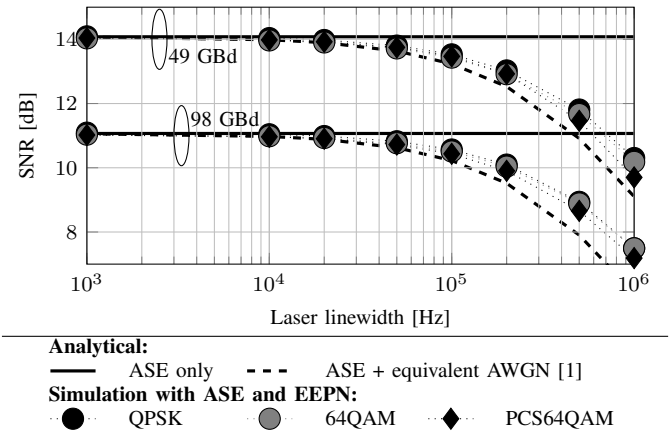

To further investigate format dependence of EEPN penalty, in presence of optimized BPS, in Fig.~\ref{SNR_LW_MF} we plot the received SNR versus LO laser linewidth for a 49~GBd and a 98~Gbd transmission using QPSK (black circles), 64QAM (gray circles) and the PCS64QAM (black diamonds), through 6600~km EX3000, neglecting nonlinearity, with OSNR equal to 20~dB/0.1nm. The solid line shows the theoretical performance in the absence of EEPN, whereas the dashed line indicates the analytical performance when EEPN is modeled as an equivalent AWGN using Eq. (3).  At low linewidths, all the formats reach the theoretical SNR without EEPN, which differs by 3~dB between 49 and 98~GBd, corresponding only to a doubling of ASE noise in the channel bandwidth. Note that, for all formats, the achieved SNR with BPS is higher than what is analytically predicted with equivalent AWGN. At high linewidths, the excess EEPN penalty PCS64QAM with respect to QPSK is inferior to 0.3~dB at 1~MHz linewidth. The takeaway is that the EEPN penalty is only mildly format dependent. For simplicity, in the rest of the paper, we will only focus on the emerging probabilistic format to study EEPN penalty on long-haul transmission system design, as it represents the worst case.

\subsection{Experimental validation}

\begin{figure}
\includegraphics[width=\linewidth]{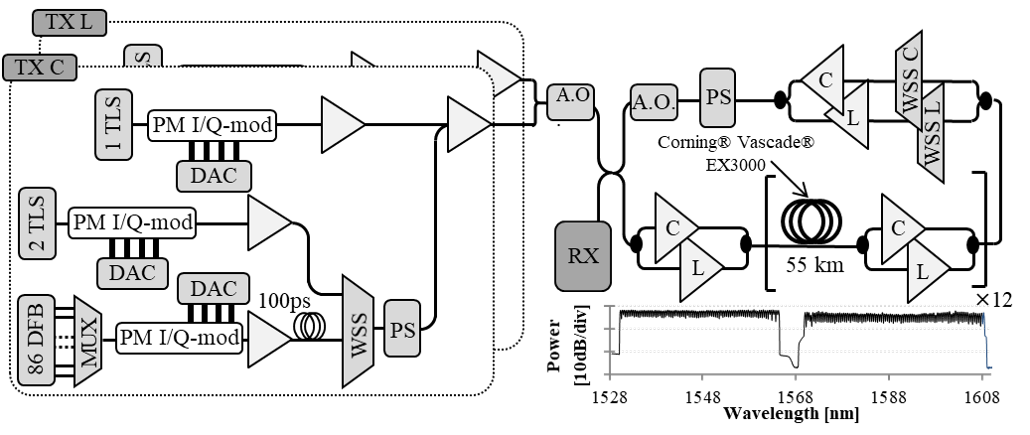}
\caption{Experimental setup for experimental WDM 49~GBd PCS64QAM transmission over 6600~km EX3000}
\label{exp_setup}
\end{figure}

To assess the impact of EEPN on BPS observed in the simulations, we performed an experimental characterization using a C+L recirculating loop depicted in Fig.~\ref{exp_setup}. The C-band transmitter (TX C) is described in details on the Figure. The test channel is a single tunable laser source (TLS) at wavelength 1545.72~nm modulated with a polarization-multiplexed I/Q (PM I/Q) modulator, surrounded by 2 loading channels, each of which driven by a distinct modulator, and further multiplexed by 86 loading channels modulated by a third dedicated modulator. Each modulator is driven by a digital-to-analogue (DAC) loaded with different randomly-generated sequences operating at 49~GBd. A wavelength selective switch (WSS) and a coupler are used to combine the measured channel with the loading channels. A fiber spool and a polarization scrambler (PS) ensure de-correlation of the loading channels. This TX C is coupled to a similar L-band transmitter (TX L) before being launched in the recirculating loop. The loop is composed of twelve 55~km spans of EX3000 fiber. The loop includes a gain equalizer WSS per band and a loop synchronous polarization scrambler. After 10 loops, the signal is sent to a coherent receiver and sampled by a real-time scope with 33~GHz bandwidth, working at 80~Gsamples/s. 

\begin{figure}
\centering
% This file was created by matplotlib2tikz v0.6.18.
\begin{tikzpicture}
\tikzstyle{every node}=[font=\figuretextsize]
\begin{axis}[
axis on top,
tick pos=both,
xlabel={CPE averaging length [symbols]},
xmajorgrids,
xmin=100, xmax=10000,
xminorgrids,
xmode=log,
ylabel={SNR [dB]},
ylabel near ticks,
xlabel near ticks,
ymajorgrids,
ymin=9, ymax=12,
yminorgrids,
width=\linewidth,
height=6cm
]

\addplot [white!50.196078431372548!black, mark=diamond*,mark size=4, mark options={solid,draw=black}, dotted, forget plot]
table [row sep=\\]{%
300	9.001772 \\
500	9.655025 \\
700	9.822213 \\
1100	9.954702 \\
1500	10.02352 \\
2300	10.082 \\
2900	10.08797 \\
3900	10.11637 \\
};

\addplot [white!50.196078431372548!black, mark=diamond*,mark size=4, mark options={solid,draw=black}, dotted, forget plot]
table [row sep=\\]{%
300	10.2165433333333 \\
500	10.7359066666667 \\
700	10.85174 \\
1100	10.9333566666667 \\
1500	10.96297 \\
2300	10.9828666666667 \\
2900	10.9868066666667 \\
3900	10.9825233333333 \\
};

\addplot [white!50.196078431372548!black, mark=diamond*,mark size=4, mark options={solid,draw=black}, dotted, forget plot]
table [row sep=\\]{%
300	11.4686775 \\
500	11.711215 \\
700	11.7713625 \\
1100	11.811865 \\
1500	11.82556 \\
2300	11.8261475 \\
2900	11.8175675 \\
3900	11.7986225 \\
};\label{SNR_CPE_B2B}
\addplot [black, mark=*, mark size=4, mark options={solid,draw=black,fill=white!50.196078431372548!black}, forget plot]
table [row sep=\\]{%
300	9.1185324 \\
500	10.34703 \\
700	10.715066 \\
1100	10.923558 \\
1500	10.989642 \\
2300	11.025954 \\
2900	11.024118 \\
3900	11.00539 \\
};\label{SNR_CPE_Pin16}
\addplot [black, mark=*, mark size=4, mark options={solid,fill=white}, forget plot]
table [row sep=\\]{%
300	9.4794612 \\
500	10.54087 \\
700	10.751336 \\
1100	10.91032 \\
1500	10.920452 \\
2300	10.889528 \\
2900	10.821778 \\
3900	10.736632 \\
};\label{SNR_CPE_Pin20}

\coordinate (opt20a) at (axis cs:1600,10.8);
\coordinate (opt20b) at (axis cs:1600,10.8);
\coordinate (opt16a) at (axis cs:2900,10.9);
\coordinate (opt16b) at (axis cs:2900,10.9);

\coordinate (leg10) at (axis cs:390,11.6);
\coordinate (leg1) at (axis cs:350,11.7);

\coordinate (leg20) at (axis cs:390,10.5);
\coordinate (leg2) at (axis cs:350, 10.6);

\coordinate (leg30) at (axis cs:390,9.4);
\coordinate (leg3) at (axis cs:350,9.5);
\end{axis}
\begin{axis}[
xshift=.45\linewidth,yshift=.7cm,width=0.5\linewidth,height=3cm,
axis on top,
tick pos=both,
xlabel={Total launched power [dBm]},
xmin=15, xmax=21,
ylabel={SNR [dB]},
ymin=10, ymax=12,
x label style={at={(axis description cs:0.5,0.15)},anchor=north},
y label style={at={(axis description cs:0.35,0.5)},anchor=south},
xmajorgrids,
ymajorgrids,
ytick={10,12},
axis background/.style={fill=white}
]
\addplot [black, mark=*, mark size=2, mark options={solid}, only marks, forget plot]
table [row sep=\\]{%
16	11.027416 \\
17	11.511908 \\
18	11.652442 \\
19	11.454168 \\
20	10.920452 \\
};
\addplot [black, mark=*, mark size=4, mark options={solid,fill=white!50.196078431372548!black}, only marks, forget plot]
table [row sep=\\]{%
16	11.027416 \\
};
\addplot [black, fill=, mark=*, mark size=4, mark options={solid,fill=white}, only marks, forget plot]
table [row sep=\\]{%
20	10.920452 \\
};
\addplot [black, forget plot]
table [row sep=\\]{%
15	10.2071655999995 \\
15.0606060606061	10.2665900841134 \\
15.1212121212121	10.3247682221693 \\
15.1818181818182	10.3817000141672 \\
15.2424242424242	10.4373854601071 \\
15.3030303030303	10.4918245599891 \\
15.3636363636364	10.545017313813 \\
15.4242424242424	10.596963721579 \\
15.4848484848485	10.647663783287 \\
15.5454545454545	10.697117498937 \\
15.6060606060606	10.7453248685291 \\
15.6666666666667	10.7922858920631 \\
15.7272727272727	10.8380005695392 \\
15.7878787878788	10.8824689009573 \\
15.8484848484848	10.9256908863174 \\
15.9090909090909	10.9676665256195 \\
15.969696969697	11.0083958188636 \\
16.030303030303	11.0478787660498 \\
16.0909090909091	11.086115367178 \\
16.1515151515152	11.1231056222481 \\
16.2121212121212	11.1588495312604 \\
16.2727272727273	11.1933470942146 \\
16.3333333333333	11.2265983111108 \\
16.3939393939394	11.2586031819491 \\
16.4545454545455	11.2893617067294 \\
16.5151515151515	11.3188738854517 \\
16.5757575757576	11.347139718116 \\
16.6363636363636	11.3741592047223 \\
16.6969696969697	11.3999323452706 \\
16.7575757575758	11.424459139761 \\
16.8181818181818	11.4477395881934 \\
16.8787878787879	11.4697736905678 \\
16.9393939393939	11.4905614468842 \\
17	11.5101028571427 \\
17.0606060606061	11.5283979213431 \\
17.1212121212121	11.5454466394856 \\
17.1818181818182	11.5612490115701 \\
17.2424242424242	11.5758050375966 \\
17.3030303030303	11.5891147175651 \\
17.3636363636364	11.6011780514756 \\
17.4242424242424	11.6119950393282 \\
17.4848484848485	11.6215656811228 \\
17.5454545454545	11.6298899768594 \\
17.6060606060606	11.636967926538 \\
17.6666666666667	11.6427995301586 \\
17.7272727272727	11.6473847877213 \\
17.7878787878788	11.6507236992259 \\
17.8484848484848	11.6528162646726 \\
17.9090909090909	11.6536624840613 \\
17.969696969697	11.653262357392 \\
18.030303030303	11.6516158846648 \\
18.0909090909091	11.6487230658795 \\
18.1515151515152	11.6445839010363 \\
18.2121212121212	11.6391983901351 \\
18.2727272727273	11.6325665331759 \\
18.3333333333333	11.6246883301587 \\
18.3939393939394	11.6155637810835 \\
18.4545454545455	11.6051928859504 \\
18.5151515151515	11.5935756447593 \\
18.5757575757576	11.5807120575102 \\
18.6363636363636	11.5666021242031 \\
18.6969696969697	11.551245844838 \\
18.7575757575758	11.534643219415 \\
18.8181818181818	11.5167942479339 \\
18.8787878787879	11.4976989303949 \\
18.9393939393939	11.4773572667979 \\
19	11.4557692571429 \\
19.0606060606061	11.4329349014299 \\
19.1212121212121	11.408854199659 \\
19.1818181818182	11.3835271518301 \\
19.2424242424242	11.3569537579431 \\
19.3030303030303	11.3291340179982 \\
19.3636363636364	11.3000679319954 \\
19.4242424242424	11.2697554999345 \\
19.4848484848485	11.2381967218157 \\
19.5454545454545	11.2053915976388 \\
19.6060606060606	11.171340127404 \\
19.6666666666667	11.1360423111112 \\
19.7272727272727	11.0994981487605 \\
19.7878787878788	11.0617076403517 \\
19.8484848484848	11.022670785885 \\
19.9090909090909	10.9823875853602 \\
19.969696969697	10.9408580387775 \\
20.030303030303	10.8980821461369 \\
20.0909090909091	10.8540599074382 \\
20.1515151515152	10.8087913226815 \\
20.2121212121212	10.7622763918669 \\
20.2727272727273	10.7145151149943 \\
20.3333333333333	10.6655074920637 \\
20.3939393939394	10.6152535230751 \\
20.4545454545455	10.5637532080285 \\
20.5151515151515	10.511006546924 \\
20.5757575757576	10.4570135397615 \\
20.6363636363636	10.401774186541 \\
20.6969696969697	10.3452884872625 \\
20.7575757575758	10.287556441926 \\
20.8181818181818	10.2285780505315 \\
20.8787878787879	10.1683533130791 \\
20.9393939393939	10.1068822295687 \\
21	10.0441648000003 \\
};
%\coordinate (nlt20) at (axis cs:19.6,11.05);
\coordinate (nlt16) at (axis cs:16.1,11.2);
\end{axis}

%\draw [->,>=stealth,black,ultra thick] (opt20)--(nlt20);
%\draw [->,>=stealth,ultra thick,white!50.196078431372548!black] (opt16)--(nlt16);
\draw[] (leg10) --  (leg1) node[above left]{OSNR = 19 dB};
\draw[] (leg20) -- (leg2) node[above left]{OSNR = 18 dB};
\draw[] (leg30) -- (leg3) node[above left]{OSNR = 17 dB};
\end{tikzpicture}

\hrule
\begin{tabular}{ll}
\ref{SNR_CPE_B2B} & Back-to-back setup (without EEPN) \\
\ref{SNR_CPE_Pin16} & Transmission (total launched power 2dB below NLT) \\
\ref{SNR_CPE_Pin20} & Transmission (total launched power 2dB over NLT) \\
\end{tabular}
\hrule
\caption{Experimental SNR vs. BPS averaging length, for a back-to-back configuration (without EEPN) and a WDM 49~GBd PCS64QAM transmission over 6600~km EX3000, for total launched power 2~dB below or over NLT. Inset: SNR vs. total launched power.}
\label{SNR_CPE_exp}
\end{figure}
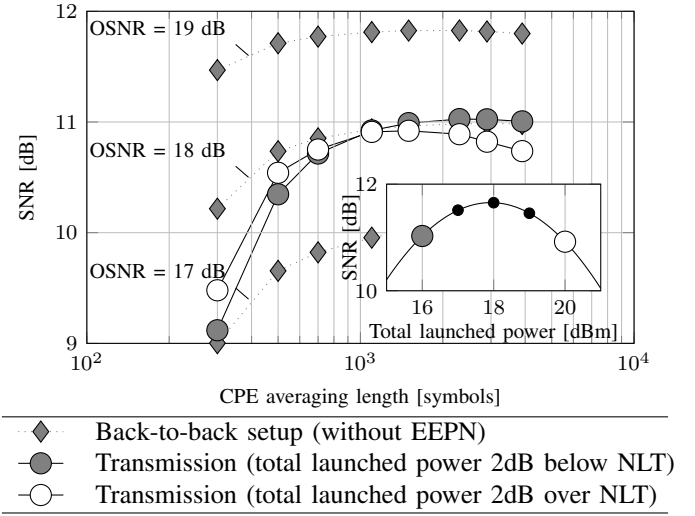

Fig.~\ref{SNR_CPE_exp} shows the SNR vs. BPS averaging length, after DSP including EDC, polarization de-multiplexing with a multi-modulus algorithm (MMA) preceded by pre-convergence by constant modulus algorithm (CMA), carrier frequency estimation and phase recovery using BPS and pilot-based cycle-slip removal using 1\% pilots and least-mean square adaptive post-equalization. The dashed lines represent three measured back-to-back SNR vs. filter length curves for OSNRs equal to  from 17~dB, 18~dB and 19~dB. The evolution of the SNR in these back-to-back measurements (hence without dispersion) is given as reference for various amount of ASE noise, as both EEPN and fiber nonlinear impairments are absent in back-to-back measurements. Then, we plot in gray circles the performance of the transmission case, after 6600~km EX3000, with a launched power of 16~dBm. This launched power has been chosen, as shown on the bell curve in the inset, to be 2~dB below the nonlinear threshold (NLT) to ensure that the nonlinear distortions are negligible, but EEPN is present. We observe that at low BPS averaging length the transmission SNR falls more sharply than back-to-back SNR, and that in the transmission case the optimal performance is achieved for longer averaging block size compared to the back-to-back case, as was observed in the numerical simulations of the previous section. We plot in white circles the transmission SNR with launched power of 20~dBm, which corresponds to 2~dB above the NLT. In the presence of fiber nonlinearity, the performance drops for long averaging block size, ensuring that the difference of behavior between the previous transmission case (at 16~dBm) and the back-to-back configuration is not attributable nonlinearities, but the signature of the presence of EEPN. As the BPS behaves differently in the transmission case and in back-to-back case with various ASE variances, this shows that EEPN must not be considered as an equivalent AWGN and the BPS must be taken into account to evaluate system performance.

\section{EEPN impact on system design}\label{sec:impact}

\subsection{OSNR penalty definition}

\begin{figure}[h!]
% This file was created by matplotlib2tikz v0.6.18.
\begin{tikzpicture}
\tikzstyle{every node}=[font=\figuretextsize]
\begin{axis}[
axis on top,
tick pos=both,
xlabel={OSNR [dB/0.1nm]},
xmajorgrids,
xmin=16, xmax=24,
xminorgrids,
ylabel={SNR [dB]},
ylabel near ticks,
xlabel near ticks,
ymajorgrids,
ymin=9, ymax=16,
yminorgrids,
width=\linewidth,
height=5cm
]
\addplot [very thick, black, forget plot]
table [row sep=\\]{%
16	10.0784 \\
18	12.0784 \\
20	14.0784 \\
22	16.0784 \\
24	18.0784 \\
26	20.0784 \\
};
\label{SNR_OSNR_ASE}

\addplot [very thick, black, dashed, forget plot]
table [row sep=\\]{%
16	9.3941 \\
18	11.0387 \\
20	12.529 \\
22	13.8268 \\
24	14.9044 \\
26	15.7541 \\
};
\label{SNR_OSNR_ASEeq}

\addplot [ lightgray!66.928104575163388!black, mark=diamond*, mark size=4, mark options={solid,draw=black}, forget plot]
table [row sep=\\]{%
16	9.35824665607369 \\
18	11.6648702819244 \\
20	13.9066544159653 \\
22	15.9825033266193 \\
24	18.0225418750453 \\
26	19.9970775087228 \\
};
\label{SNR_OSNR_D0_BPS}

%\addplot [dotted, black, mark=*, mark size=4, mark options={solid,fill=white}, only marks, forget plot]
%table [row sep=\\]{%
%16	9.41122695410258 \\
%18	11.0674752786201 \\
%20	12.5731173340865 \\
%22	13.8895158317294 \\
%24	14.9878743020268 \\
%26	15.8583364503569 \\
%};
%\label{SNR_OSNR_D1_PT}

%\addplot [dotted, black, mark=*, mark size=4, mark options={solid}, forget plot]
%table [row sep=\\]{%
%16	9.65134633471099 \\
%18	11.4409725494808 \\
%20	13.0865547580324 \\
%22	14.5785493945299 \\
%24	15.8789374217451 \\
%26	16.9601907026497 \\
%};
%\label{SNR_OSNR_D1_IDR}

\addplot [ lightgray!66.928104575163388!black, mark=*, mark size=4, mark options={solid,draw=black}, forget plot]
table [row sep=\\]{%
16	9.19969201773413 \\
18	11.1543560376798 \\
20	12.9237095585423 \\
22	14.4385812968023 \\
24	15.740168674812 \\
26	16.812140541501 \\
};
\label{SNR_OSNR_D1_BPS}

\coordinate (a) at (axis cs: 16.1,19.9);

\coordinate (b0) at (axis cs: 17.92,12);
\coordinate (b1) at (axis cs: 17.92,14.2);
\coordinate (b2) at (axis cs: 17.92,9.8);

\coordinate (c0) at (axis cs: 18.95,12);
\coordinate (c1) at (axis cs: 18.95,14.2);
\coordinate (c2) at (axis cs: 18.95,11);

\coordinate (d0) at (axis cs:  19.29,12);
\coordinate (d2) at (axis cs:  19.29,11);

\coordinate (e0) at (axis cs:  18.3,12);
\coordinate (e2) at (axis cs:  18.3,9.8);
\end{axis}

\draw[->,>=stealth,black,ultra thick] (b1) -- (c1) node[midway,above]{\begin{tabular}{l}Total OSNR penalty \\ 1~dB for $SNR_{ref}$~=~12~dB \end{tabular}};
%\draw[->,>=stealth,black,ultra thick] (b0) -- (c0);

\draw[black, thick] (b1)--(b0);
\draw[black, thick] (c1)--(c0);

\draw[->,>=stealth,black,ultra thick] (c2) -- (d2) node[right]{\begin{tabular}{l} Penalty overestimation \\ 0.3~dB for $SNR_{ref}$~=~12~dB \end{tabular}};
%\draw[->,>=stealth,black,ultra thick] (c0) -- (d0);
\draw[black, thick] (c2)--(c0);
\draw[black, thick] (d2)--(d0);

\draw[->,>=stealth,black,ultra thick] (b2) -- (e2) node[right]{\begin{tabular}{l} PN Penalty \\ 0.3~dB for $SNR_{ref}$~=~12~dB \end{tabular}};
%\draw[->,>=stealth,black,ultra thick] (c0) -- (d0);
\draw[black, thick] (b2)--(b0);
\draw[black, thick] (e2)--(e0);

\end{tikzpicture}

\centering
\hrule
{\figuretextsize
\begin{tabular}{ll}
\multicolumn{2}{l}{\bf Analytical:} \\
\ref{SNR_OSNR_ASE}  ASE only & \ref{SNR_OSNR_ASEeq}  ASE + equivalent AWGN \cite{shieh2008}\\
\multicolumn{2}{l}{\bf Simulation with optimized BPS:} \\
\ref{SNR_OSNR_D0_BPS}  with ASE, without EEPN &  \ref{SNR_OSNR_D1_BPS}  with ASE and EEPN  \\

\end{tabular}
}
\hrule
\caption{SNR vs. OSNR (referred to 0.1 nm) of a single-channel single-polarization 49 GBd PCS64QAM through 6600~km EX3000, neglecting nonlinearity, and LO linewidth equal to 200~kHz}
\label{pyOSNR_SNR}
\end{figure}
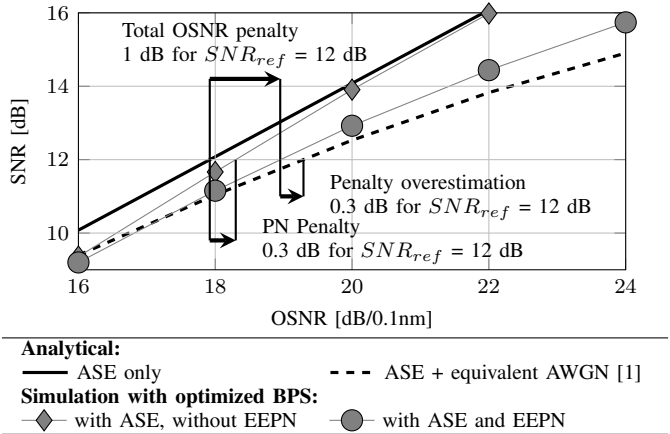

To assess the amount by which system performance could be underestimated due to simplifying assumptions, in Fig.~\ref{pyOSNR_SNR} we plot the estimated SNR after digital signal processing vs. OSNR curves, assuming the same numerical setup as in Fig.~\ref{pyPDF} and sweeping the ASE noise variance. The operative SNR at the considered configuration was set to $SNR_{ref}~=~12$~dB according to\cite{amir2017} and the EEPN impact is considered in term of OSNR penalty to reach this operative SNR. The solid line is the theoretical SNR vs. OSNR curve assuming both laser linewidth and dispersion are zero. Next, we turn on LO phase noise but keep dispersion to zero: gray diamonds correspond to BPS, where the averaging length has been optimized to maximize the SNR. The penalty between solid black line and gray diamonds corresponds to the residual noise coming from the use of BPS as CPR and is denoted PN penalty. Next, we turn on dispersion (hence EEPN) and optimize BPS, to obtain gray circles, which represent the true system performance including EEPN and impact of DSP. Performance degradation from gray diamonds to gray circles is the EEPN penalty, whereas the degradation from black solid line to gray circles represents the total OSNR penalty due to both PN and EEPN, which is about 1~dB for this specific configuration and operative SNR of 12~dB. For low OSNRs the BPS performance without EEPN (diamonds) converges to that of BPS performance with EEPN (circles), showing that in this regime the performance is mainly due ASE and PN, and not EEPN. The dashed line is the analytical performance estimated when EEPN is modeled as an independent AWGN with the variance computed in \cite{shieh2008}, and shows that ignoring DSP overestimates the EEPN penalty by 0.3~dB in this configuration.

\subsection{Impact on modern and future transoceanic systems}
\def\fig8size{4.5cm}

\def\figwidth{0.32\linewidth}
\begin{figure*}[t!]

\begin{tabular}{ccc}
(a) & (b) & (c) \\
% This file was created by matplotlib2tikz v0.6.18.
\begin{tikzpicture}
\tikzstyle{every node}=[font=\figuretextsize]
\begin{axis}[
axis on top,
tick pos=both,
xlabel={Laser linewidth [Hz]},
xmajorgrids,
xmin=1000, xmax=500000,
xminorgrids,
xmode=log,
ylabel={OSNR penalty [dB]},
ylabel near ticks,
xlabel near ticks,
ymajorgrids,
ymin=0, ymax=2,
yminorgrids,
height=\fig8size,
width=\figwidth
]
\addplot [very thick, black, dashed, forget plot]
table [row sep=\\]{%
1000	0.00582016217839865 \\
10000	0.0580826038613189 \\
20000	0.117187753413322 \\
50000	0.301840638401824 \\
100000	0.631840702781847 \\
200000	1.36847582365967 \\
500000	4.80769387819008 \\
};\label{OSNRp49_th}

\addplot [black, mark=diamond*, mark size=4, mark options={solid,draw=black,fill=white}, forget plot]
table [row sep=\\]{%
1000	0.0395621986110299 \\
10000	0.0743920878380457 \\
20000	0.0987108332103901 \\
50000	0.159331429282958 \\
100000	0.239678097094846 \\
200000	0.377384824619597 \\
500000	0.702464036159174 \\
};\label{OSNRp49_PN}

\addplot [black, mark=diamond*, mark size=4, mark options={solid,draw=black,fill=black}, forget plot]
table [row sep=\\]{%
1000	0.0452917741809316 \\
10000	0.0993098675626527 \\
20000	0.13793485121635 \\
50000	0.233942791779096 \\
100000	0.367160272524096 \\
200000	0.605547976243237 \\
500000	1.04889952061439 \\
};\label{OSNRp49_PN_TXLO}

\addplot [black, mark=*, mark size=4, mark options={solid,draw=black, fill=white}, forget plot]
table [row sep=\\]{%
1000	0.0394754741169514 \\
10000	0.0850988130826913 \\
20000	0.13114422971211 \\
50000	0.267201608935224 \\
100000	0.512056065026012 \\
200000	1.03427902852559 \\
500000	3.03748220621301 \\
};\label{OSNRp49_EEPN}

\addplot [black, mark=*, mark size=4, mark options={solid,draw=black,fill=black}, forget plot]
table [row sep=\\]{%
1000	0.0457222136332298 \\
10000	0.112485204460693 \\
20000	0.177763807778401 \\
50000	0.373293706160304 \\
100000	0.705035194283546 \\
200000	1.39560983867541 \\
500000	4.26637212027864 \\
};\label{OSNRp49_EEPN_TXLO}

\coordinate (leg1) at (axis cs:1100,1.95);
\coordinate (leg2) at (axis cs:4e4,0.95);

\coordinate (a0) at (axis cs:2e5,1.03);
\coordinate (b0) at (axis cs:2e5,1.37);

\coordinate (a1) at (axis cs:3e5,1.03);
\coordinate (b1) at (axis cs:3e5,1.37);
\end{axis}

%\draw[] (a0)--(a1);
%\draw[] (b0)--(b1);
%\draw[->,>=stealth,black,ultra thick] (a1) -- (b1)node[midway,right]{0.3~dB};

\draw[fill=white] (leg1) rectangle node[black]{
\begin{tabular}{l}
PCS64QAM \\
6600~km EX3000 \\
49~GBd \\
$SNR_{ref}$ = 12 dB
\end{tabular}
} (leg2);

\end{tikzpicture} & 
% This file was created by matplotlib2tikz v0.6.18.
\begin{tikzpicture}
\tikzstyle{every node}=[font=\figuretextsize]
\begin{axis}[
axis on top,
tick pos=both,
xlabel={Laser linewidth [Hz]},
xmajorgrids,
xmin=1000, xmax=500000,
xminorgrids,
xmode=log,
ylabel={OSNR penalty [dB]},
ylabel near ticks,
xlabel near ticks,
ymajorgrids,
ymin=0, ymax=2,
yminorgrids,
height=\fig8size,
width=\figwidth
]
\addplot [very thick, black, dashed, forget plot]
table [row sep=\\]{%
1000	0.0118481199237905 \\
10000	0.119974678993319 \\
20000	0.242920750738083 \\
50000	0.628754248069058 \\
100000	1.35679515506645 \\
200000	3.31736444627515 \\
500000	nan \\
};
\addplot [black, mark=diamond*, mark size=4, mark options={solid,draw=black,fill=white}, forget plot]
table [row sep=\\]{%
1000	0.0426628943102507 \\
10000	0.0628476036230197 \\
20000	0.0822654718591345 \\
50000	0.124633617758668 \\
100000	0.17568058455241 \\
200000	0.25370709927002 \\
500000	0.438763326461931 \\
};
\addplot [black, mark=*, mark size=4, mark options={solid,draw=black,fill=white}, forget plot]
table [row sep=\\]{%
1000	0.0443692942929381 \\
10000	0.111973939677068 \\
20000	0.189530353367321 \\
50000	0.427916125120007 \\
100000	0.843301373723317 \\
200000	1.83185084500382 \\
500000	nan \\
};

\addplot [black, mark=diamond*, mark size=4, mark options={solid,draw=black}, forget plot]
table [row sep=\\]{%
1000	0.043694563501159 \\
10000	0.0801872895923061 \\
20000	0.108346432619772 \\
50000	0.174099389436041 \\
100000	0.249802517023074 \\
200000	0.371983535592502 \\
500000	0.687438340809763 \\
};
\addplot [black, mark=*, mark size=4, mark options={solid,draw=black}, forget plot]
table [row sep=\\]{%
1000	0.0484897546569947 \\
10000	0.138512950229039 \\
20000	0.232116235517807 \\
50000	0.513199735328062 \\
100000	0.980960741055856 \\
200000	2.17236016943341 \\
500000	nan \\
};

\coordinate (leg1) at (axis cs:1100,1.95);
\coordinate (leg2) at (axis cs:4e4,0.95);

\coordinate (a0) at (axis cs:1e5,0.84);
\coordinate (b0) at (axis cs:1e5,1.35);

\coordinate (a1) at (axis cs:2e5,0.84);
\coordinate (b1) at (axis cs:2e5,1.35);
\end{axis}

%\draw[] (a0)--(a1);
%\draw[] (b0)--(b1);
%\draw[->,>=stealth,black,ultra thick] (a1) -- (b1)node[midway,right]{0.5~dB};

\draw[fill=white] (leg1) rectangle node[black]{
\begin{tabular}{l}
PCS64QAM \\
6600~km EX3000 \\
98~GBd \\
$SNR_{ref}$ = 12 dB
\end{tabular}
} (leg2);

\end{tikzpicture} & 
% This file was created by matplotlib2tikz v0.6.18.
\begin{tikzpicture}
\tikzstyle{every node}=[font=\figuretextsize]
\begin{axis}[
axis on top,
tick pos=both,
xlabel={Laser linewidth [Hz]},
xmajorgrids,
xmin=1000, xmax=500000,
xminorgrids,
xmode=log,
ylabel={OSNR penalty error [dB]},
ymajorgrids,
ymin=-0.5, ymax=1.5,
yminorgrids,
width=\figwidth,
height=\fig8size
]

%\addplot [black, mark=*, mark size=4, mark options={solid,fill=black}, forget plot]
%table [row sep=\\]{%
%1000	-0.0336553119385528 \\
%10000	-0.0270162092213724 \\
%20000	-0.0139564762987874 \\
%50000	0.0346390294666001 \\
%100000	0.119784637755835 \\
%200000	0.334196795134076 \\
%500000	1.77021167197707 \\
%};\label{OSNRmis_49_6600}

\addplot [black, mark=*, mark size=4, mark options={solid,draw=black,fill=black}, forget plot]
table [row sep=\\]{%
1000	-0.0399020514548312 \\
10000	-0.0544026005993743 \\
20000	-0.0605760543650788 \\
50000	-0.0714530677584797 \\
100000	-0.0731944915016989 \\
200000	-0.02713401501574 \\
500000	0.54132175791144 \\
};\label{OSNRmis_49_6600_TXLO}

%\addplot [black, mark=*, mark size=4, mark options={solid,fill=black}, forget plot]
%table [row sep=\\]{%
%1000	-0.0325211743691476 \\
%10000	0.00800073931625178 \\
%20000	0.0533903973707623 \\
%50000	0.200838122949051 \\
%100000	0.513493781343136 \\
%200000	1.48551360127133 \\
%500000	nan \\
%};\label{OSNRmis_98_6600}

\addplot [black, mark=*, mark size=4, mark options={solid,draw=black,fill=black}, forget plot]
table [row sep=\\]{%
1000	-0.0366416347332041 \\
10000	-0.0185382712357196 \\
20000	0.0108045152202756 \\
50000	0.115554512740996 \\
100000	0.375834414010598 \\
200000	1.14500427684174 \\
500000	nan \\
};\label{OSNRmis_98_6600_TXLO}

\coordinate (leg1) at (axis cs:1100,1.45);
\coordinate (leg2) at (axis cs:4e4,0.7);

\coordinate (a) at (axis cs: 1e5,0.6);
\coordinate (d) at (axis cs: 3e4,0.5);
\coordinate (b) at (axis cs: 1.5e5,0);
\coordinate (c) at (axis cs: 1.5e5,-0.3);

\end{axis}

\draw (a) ellipse(0.3cm and 0.1cm);
\draw (d) node{98 GBd};
\draw (b) ellipse(0.1cm and 0.3cm) ;
\draw (c) node{49 GBd};

\draw[fill=white] (leg1) rectangle node[black]{
\begin{tabular}{l}
PCS64QAM \\
6600km EX3000 \\
$SNR_{ref}=12$ dB \\
\end{tabular}
}(leg2);

\end{tikzpicture} \\
(d) & (e) & (f) \\
% This file was created by matplotlib2tikz v0.6.18.
\begin{tikzpicture}
\tikzstyle{every node}=[font=\figuretextsize]
\begin{axis}[
axis on top,
tick pos=both,
xlabel={Laser linewidth [Hz]},
xmajorgrids,
xmin=1000, xmax=500000,
xminorgrids,
xmode=log,
ylabel={OSNR penalty [dB]},
ylabel near ticks,
xlabel near ticks,
ymajorgrids,
ymin=0, ymax=2,
yminorgrids,
height=\fig8size,
width=\figwidth
]

\addplot [very thick, black, dashed, forget plot]
table [row sep=\\]{%
1000	0.00455222939498334 \\
10000	0.0461494199535952 \\
20000	0.092620300178071 \\
50000	0.235081460884443 \\
100000	0.488939201730911 \\
200000	1.04853152854041 \\
500000	3.29564714987972 \\
};

\addplot [black, mark=diamond*, mark size=4, mark options={solid,draw=black,fill=white}, forget plot]
table [row sep=\\]{%
1000	0.0487722062607236 \\
10000	0.071649633143327 \\
20000	0.0909520850072667 \\
50000	0.143384287261123 \\
100000	0.228756759522261 \\
200000	0.354857941765971 \\
300000	0.470977895354766 \\
500000	0.675082490428249 \\
};

\addplot [black, mark=*, mark size=4, mark options={solid,draw=black,fill=white}, forget plot]
table [row sep=\\]{%
1000	0.0430600555094944 \\
10000	0.0748183756141998 \\
20000	0.109571395762453 \\
50000	0.20534759384633 \\
100000	0.397554104081959 \\
200000	0.827559370362172 \\
300000	1.27200415700903 \\
500000	2.20807283719088 \\
};

\addplot [black, mark=diamond*, mark size=4, mark options={solid,draw=black}, forget plot]
table [row sep=\\]{%
1000	0.0515816392412471 \\
10000	0.0916616684911453 \\
20000	0.124411024171714 \\
50000	0.217811346685076 \\
100000	0.357060249773875 \\
200000	0.572184637402076 \\
300000	0.736847637882004 \\
500000	0.945580600838934 \\
};

\addplot [black, mark=*, mark size=4, mark options={solid,draw=black}, forget plot]
table [row sep=\\]{%
1000	0.047240701646432 \\
10000	0.0991721446011393 \\
20000	0.148664596483783 \\
50000	0.311845392723498 \\
100000	0.59894341171789 \\
200000	1.16682581625521 \\
300000	1.73521897681044 \\
500000	3.11280556519189 \\
};

\coordinate (leg1) at (axis cs:1100,1.95);
\coordinate (leg2) at (axis cs:4e4,0.95);

\coordinate (a0) at (axis cs:1e5,0.94);
\coordinate (b0) at (axis cs:1e5,1.4);

\coordinate (a1) at (axis cs:2e5,0.94);
\coordinate (b1) at (axis cs:2e5,1.4);
\end{axis}

%\draw[] (a0)--(a1);
%\draw[] (b0)--(b1);
%\draw[->,>=stealth,black,ultra thick] (a1) -- (b1)node[midway,right]{0.46~dB};

\draw[fill=white] (leg1) rectangle node[black]{
\begin{tabular}{l}
TPCS64QAM \\ 
13419~km EX3000 \\
49~GBd \\
$SNR_{ref}$ = 7.9 dB
\end{tabular}
} (leg2);

\end{tikzpicture} & 
% This file was created by matplotlib2tikz v0.6.18.
\begin{tikzpicture}
\tikzstyle{every node}=[font=\figuretextsize]
\begin{axis}[
axis on top,
tick pos=both,
xlabel={Laser linewidth [Hz]},
xmajorgrids,
xmin=1000, xmax=500000,
xminorgrids,
xmode=log,
ylabel={OSNR penalty [dB]},
ylabel near ticks,
xlabel near ticks,
ymajorgrids,
ymin=0, ymax=2,
yminorgrids,
height=\fig8size,
width=\figwidth
]
\addplot [very thick, black, dashed, forget plot]
table [row sep=\\]{%
1000	0.00935075165363841 \\
10000	0.09466753027748 \\
20000	0.191278116517019 \\
50000	0.492553324453322 \\
100000	1.03015634833517 \\
200000	2.41280787468247 \\
300000	4.39177551856256 \\
500000	nan \\
};
\addplot [black, mark=diamond*, mark size=4, mark options={solid,fill=white}, forget plot]
table [row sep=\\]{%
1000	0.0504645327242024 \\
10000	0.063011509948506 \\
20000	0.0741997781259762 \\
50000	0.108397214658179 \\
100000	0.158578622630341 \\
200000	0.253213826030233 \\
300000	0.326798606693426 \\
500000	0.442562530135483 \\
};
\addplot [black, mark=*, mark size=4, mark options={solid,fill=white}, forget plot]
table [row sep=\\]{%
1000	0.0489428000791783 \\
10000	0.101248436755192 \\
20000	0.158456108741948 \\
50000	0.337327413849746 \\
100000	0.630048968476849 \\
200000	1.24891366630046 \\
300000	2.075620287603 \\
500000	4.08089951968085 \\
};

\addplot [black, mark=diamond*, mark size=4, mark options={solid}, forget plot]
table [row sep=\\]{%
1000	0.0511618417506376 \\
10000	0.0751179355260447 \\
20000	0.0987384746544961 \\
50000	0.160457171123124 \\
100000	0.246621834124614 \\
200000	0.379284854038492 \\
300000	0.484613096841009 \\
500000	0.647312119201963 \\
};
\addplot [black, mark=*, mark size=4, mark options={solid}, forget plot]
table [row sep=\\]{%
1000	0.0500955497226769 \\
10000	0.114050044567321 \\
20000	0.188240595473225 \\
50000	0.408631027718179 \\
100000	0.762909634662368 \\
200000	1.57750547982807 \\
300000	2.61860839616896 \\
500000	5.26400308106869 \\
};
\path [draw=black, fill opacity=0] (axis cs:1000,1)
--(axis cs:1000000,1);

\path [draw=black, fill opacity=0] (axis cs:1,0)
--(axis cs:1,4);

\path [draw=black, fill opacity=0] (axis cs:0,0)
--(axis cs:0,4);

\path [draw=black, fill opacity=0] (axis cs:1000,0)
--(axis cs:1000000,0);

\coordinate (leg1) at (axis cs:1100,1.95);
\coordinate (leg2) at (axis cs:4e4,0.95);

\end{axis}

\draw[fill=white] (leg1) rectangle node[black]{
\begin{tabular}{l}
TPCS64QAM \\
13419~km EX3000 \\
98~GBd \\
$SNR_{ref}$ = 7.9 dB
\end{tabular}
} (leg2);

\end{tikzpicture} & 
% This file was created by matplotlib2tikz v0.6.18.
\begin{tikzpicture}
\tikzstyle{every node}=[font=\figuretextsize]
\begin{axis}[
axis on top,
tick pos=both,
xlabel={Laser linewidth [Hz]},
xmajorgrids,
xmin=1000, xmax=500000,
xminorgrids,
xmode=log,
ylabel={OSNR penalty error [dB]},
ymajorgrids,
ymin=-0.5, ymax=1.5,
yminorgrids,
width=\figwidth,
height=\fig8size
]

%\addplot [black, mark=*, mark size=4, mark options={solid,fill=black}, forget plot]
%table [row sep=\\]{%
%1000	-0.0385078261145111 \\
%10000	-0.0286689556606046 \\
%20000	-0.0169510955843819 \\
%50000	0.0297338670381126 \\
%100000	0.0913850976489528 \\
%200000	0.220972158178242 \\
%300000	0.399002846882023 \\
%500000	1.08757431268884 \\
%700000	2.34585651261316 \\
%};\label{OSNRmis_49_13419}

\addplot [black, mark=*, mark size=4, mark options={solid,draw=black,fill=black}, forget plot]
table [row sep=\\]{%
1000	-0.0426884722514487 \\
10000	-0.0530227246475441 \\
20000	-0.0560442963057124 \\
50000	-0.0767639318390554 \\
100000	-0.110004209986979 \\
200000	-0.118294287714798 \\
300000	-0.0642119729193933 \\
500000	0.182841584687829 \\
};
\label{OSNRmis_49_13419_TXLO}

%\addplot [black, mark=*, mark size=4, mark options={solid,fill=black}, forget plot]
%table [row sep=\\]{%
%1000	-0.0395920484255399 \\
%10000	-0.00658090647771203 \\
%20000	0.0328220077750707 \\
%50000	0.155225910603576 \\
%100000	0.400107379858316 \\
%200000	1.16389420838201 \\
%500000	nan \\
%700000	nan \\
%};\label{OSNRmis_98_13419}

\addplot [black, mark=*, mark size=4, mark options={solid,fill=black}, forget plot]
table [row sep=\\]{%
1000	-0.0407447980690385 \\
10000	-0.0193825142898412 \\
20000	0.00303752104379384 \\
50000	0.0839222967351425 \\
100000	0.267246713672797 \\
200000	0.835302394854399 \\
300000	1.77316712239361 \\
500000	nan \\
};\label{OSNRmis_98_13419_TXLO}

\coordinate (leg1) at (axis cs:1100,1.45);
\coordinate (leg2) at (axis cs:4e4,0.7);

\coordinate (a) at (axis cs: 1.3e5,0.6);
\coordinate (b) at (axis cs: 1.5e5,-0.1);
\coordinate (c) at (axis cs: 2.5e5,-0.4);
\coordinate (d) at (axis cs: 3e4,0.5);
\end{axis}

\draw (a) ellipse(0.3cm and 0.1cm) ;
\draw (d) node{98 GBd};
\draw (b) ellipse(0.1cm and 0.3cm) ;
\draw (c) node{49 GBd};

\draw[fill=white] (leg1) rectangle node[black]{
\begin{tabular}{l}
TPCS64QAM \\
13419km EX3000 \\
$SNR_{ref}=7.9$ dB
\end{tabular}
}(leg2);

\end{tikzpicture} \\
\end{tabular}

\begin{center}
\hrule
{\figuretextsize
\begin{tabular}{ll|ll}
{\bf Analytical:} \\
 \multicolumn{4}{l}{\ref{OSNRp49_th}  ASE + equivalent AWGN \cite{shieh2008,kakkar2015}} \\
\multicolumn{2}{l|}{\bf Simulation with ASE and EEPN, phase noise generated by LO only:} & \multicolumn{2}{l}{\bf Simulation with ASE and EEPN, phase noise generated by LO and TX:} \\
\ref{OSNRp49_EEPN} Total OSNR penalty & \ref{OSNRp49_PN} PN penalty (no dispersion) & \ref{OSNRp49_EEPN_TXLO} Total OSNR penalty & \ref{OSNRp49_PN_TXLO} PN penalty (no dispersion)  \\
\end{tabular}
}
\hrule
\end{center}

\caption{Simulation results of ONSR penalty induced by EEPN as a function of LO linewidth for a single-channel transmission through EX3000 fiber, neglecting nonlinearity. For a transatlantic 6600~km with PCS64QAM transmission, OSNR penalty is given for 49~GBd (a) and 98~GBd (b), and OSNR penalty error between simulations and analytical model is shown in (c). For a transpacific 13419~km with TPCS64QAM transmission, (c) and (d) show OSNR penalty for 49~GBd and 98~GBd, and (e) gives the OSNR penalty error between simulations and anaylical model.}
\label{pyOSNRpenalty}
\end{figure*}

For the PCS64QAM transmission over transoceanic links of 6600 or 13419~km, with baud-rates of 49 or 98~GBd, we computed the previously defined PN and total OSNR penalties and we plot in Fig.~\ref{pyOSNRpenalty} these penalties vs. laser linewidth. Firstly, we set TX laser linewidth to 0, and only swept the LO linewidth. In this case, white circles stand for total EEPN penalty, whereas white diamonds correspond penalty only due to PN, \emph{i.e.}, we artificially set fiber dispersion coefficient to zero, so the EEPN was also forced to zero. Secondly, we performed simulations where both TX laser source and LO linewidth were swept, but for simplicity assumed these two lasers have the same linewidth. In this case black circles represent the total EEPN penalty, and black diamonds correspond to PN only penalty. For each simulation point, the BPS was separately optimized for maximum SNR. The dashed line represents the DSP-agnostic penalty predicted by the analytical model \cite{shieh2008} where EEPN is modeled as an equivalent AWGN. Given this analytical model does not capture the CPR, its predictions do not depend on TX laser linewidth 

Fig.~\ref{pyOSNRpenalty}a presents simulation results for transatlantic transmission of 49~GBd PCS64QAM over 6600~km, with a target SNR of 12~dB (parameters set according to \cite{amir2017}). For low linewidths, EEPN and PN penalties are similar and they are slightly higher than the analytical prediction, due to BPS implementation penalty. As the linewidth increases, EEPN increasingly dominates PN. When TX laser linewidth is set to zero, for LO linewidths larger than 20~kHz, the total OSNR penalty (white circles) is always lower than the pessimistic analytical prediction. When TX laser linewidth is set to that of LO, even if TX laser phase noise does not directly contribute to EEPN generation, this additional phase noise impacts BPS optimum point. As a result, the PN penalty (black diamonds) and the total penalty (black circles) increase and to keep EEPN penalty under 1~dB of OSNR, we must limit the linewidth at 130~kHz. 

Since the industry trend is towards increasing the per-channel baud rate, hence reducing transponder count and cost-per-bit, on Fig.~\ref{pyOSNRpenalty}b, we examine the OSNR vs. linewidth penalty of 98~GBd PCS64QAM on the same link as in Fig.~\ref{pyOSNRpenalty}a. In the analytical model, EEPN variance scales linearly with the baudrate: as a result, OSNR penalty increases as we double the baudrate. To keep EEPN penalty under 1~dB of OSNR, the analytical model gives a linewidth constraint on LO and TX of 70~kHz. When TX linewidth is set to 0, the total OSNR penalty is smaller than the analytical model. Besides, increasing the baudrate reduces the share of PN penalty in the total OSNR penalty. As a result, when both TX and LO linewidths are nonzero, the linewidth constraint on both lasers to keep OSNR penalty under 1~dB is loosened to 100~kHz. 

Then, we plot the OSNR penalty discrepancy between the analytical model and our DSP-aware simulations in Fig. \ref{pyOSNRpenalty}c for the case where TX laser linewidth equals LO linewidth. At 49~GBd, the analytical model overestimates by 0.5~dB the OSNR penalty at 500~kHz linewidth. At 98 GBd, the PN penalty is reduced.

As the EEPN penalty scales up with total transmission distance, we also studied a transpacific transmission of TPCS64QAM over 13419~km EX3000. In order to meet transmission experiment parameters from \cite{omar2018}, the target SNR is set to 7.9~dB, hence both EEPN and ASE increase. In Fig.~\ref{pyOSNRpenalty}d, we show simulation results from 49~GBd transmission. The PN penalty (white diamonds:TX source linewidth set to zero, black diamonds: TX laser linewidth equal to LO linewidth) is comparable to that of 6600~km transmission of Fig.~\ref{pyOSNRpenalty}a. To keep EEPN penalty under 1~dB of OSNR, we must limit TX and laser to 150~kHz.

The 98~GBd transpacific results are plotted in Fig.~\ref{pyOSNRpenalty}d. In order to keep EEPN OSNR penalty under 1~dB, the TX and LO linewidth should be kept below 120~kHz. 

Finally, we plot OSNR penalty discrepancy between the analytical model and our simulations in Fig.~\ref{pyOSNRpenalty}e. For 49~GBd transmission, the overestimation by the analytical model remains under 0.3~dB for both LO and TX linewidths smaller than 500~kHz. However, for 98~GBd transmission, we observe more than 1~dB penalty overestimation for linewidth over 220~kHz when both TX and LO generate phase noise. Please note that the revised assessment of EEPN, is not only useful for setting requirements on laser specifications, but also to separate the extra penalty due to EEPN from other sources of penalty in advanced characterizations of submarine systems, which require an accuracy of less than 0.1 dB on SNR.

\section{Conclusions}

We characterized the electronically enhanced phase noise (EEPN) in coherent receivers, taking into account the blind phase search algorithm. We showed the influence of EEPN on the behavior of this carrier phase recovery process, and quantified EEPN penalty as a function of local oscillator linewidth for emerging transoceanic systems with shaped constellations, highlighting the importance of DSP-aware performance analysis. Not taking into account the EEPN characteristics and the BPS effect on OSNR penalty can lead to overestimate its impact on system design. As both TX and LO play a role in the CPR algorithm, even if TX does not generate EEPN, both of them must be included in the simulations to estimate the full PN impact, which accounts for BPS residual error and EEPN. For high baudrate transmission, EEPN penalty becomes dominant and the overestimation is significant for current laser source performance. These penalty curves might serve as guidelines for specifications of the laser sources for high baudrate coherent systems in future.

\end{document}